\documentclass[a4paper,11pt]{article}

\usepackage{jcappub} 
\usepackage{mathtools}
\usepackage{hyperref}
\usepackage[T1]{fontenc} 
\usepackage{subcaption}
\usepackage{amsmath} 
\usepackage{amsfonts} 
\usepackage{amssymb} 
\usepackage{graphicx}
\usepackage{bm} 

\def\HI{H{\sc i}\, }

\title{\boldmath Calibration requirements for Epoch of Reionization 21-cm signal observations - V. Bias and variance in the cylindrical power spectrum from uGMRT Band~3 observations of the ELAIS-N1 field}

\def \iitbhu {Department of Physics, Indian Institute of Technology (Banaras Hindu University), Varanasi - 221005, India}
\def \iitm {Centre for Strings, Gravitation and Cosmology, Department of Physics, Indian Institute of Technology, Madras, Chennai 600036, India}
\def \iitkgp {Department of Physics, Indian Institute of Technology, Kharagpur - 721302, India}
\def \bbc {Department of Physics, Banwarilal Bhalotia College, Asansol -713303, India}
\def \rri {Raman Research Institute, Bangalore - 560080, India}
\def \lmu {University Observatory, Faculty of Physics, Ludwig Maximilian University of Munich (LMU), Scheinerstr. 1, 81679 München, Germany}
\def \iisc {Department of Physics, Indian Institute of Science, Bangalore 560012, India}
\def \ju {Department of Physics, Jadavpur University, Kolkata - 700032, India}
\def \knpg  {Department of Physics, Government Degree College, Manikpur, Chitrakoot - 210208, India}
\def \iiti {Department of Astronomy, Astrophysics and Space Engineering (DAASE), Indian Institute of Technology Indore (IIT Indore),  Indore - 453552, India}
\def \ut {Dunlap Institute for Astronomy \& Astrophysics, University of Toronto, 50 St. George Street, Toronto, ON M5S 3H4, Canada}

\author[a]{Saikat Gayen}
\author[c]{Suvam Tripathy}   
\author[e]{Saqueb Tanveer}  
\author[a]{Prasun Dutta}
\author[h]{Abhik Ghosh}      
\author[i]{Khandakar Md Asif Elahi} 
\author[j]{Sarvesh Mangla} 
\author[d]{Nirupam Roy} 
\author[c]{Samir Choudhuri} 
\author[e]{Somnath Bharadwaj} 
\author[g]{Sk Saiyad Ali}
\author[b]{Jais Kumar}
\author[d]{Srijita Pal} 
\author[k]{Arnab Chaktaborty} 
\author[f]{Rashmi Sagar} 
\author[f]{Abhirup Datta} 

\affiliation[a]{\iitbhu}
\affiliation[b]{\knpg}
\affiliation[c]{\iitm}
\affiliation[d]{\iisc}
\affiliation[e]{\iitkgp}
\affiliation[f]{\iiti}
\affiliation[g]{\ju}
\affiliation[h]{\bbc}
\affiliation[i]{\rri}
\affiliation[j]{\lmu}
\affiliation[k]{\ut}

\emailAdd{saikatgayen.rs.phy22@itbhu.ac.in}

\abstract{

Detection of the redshifted 21-cm signal from neutral hydrogen requires precise control of systematic effects arising from instrumental calibration. In this work, we investigate the impact of both time- and frequency-dependent residual gain and bandpass errors on the estimation of the multifrequency angular power spectrum (MAPS) using radio interferometric observations. 
We discuss our previously developed analytical framework for estimating bias and excess variance in 21-cm signal MAPS in the presence of these residual errors and apply it to uGMRT Band-3 observations. The statistical properties of the residual gains are derived from the data themselves, including their temporal and spectral correlations. We introduce a robust technique for  additionally flagging poor-quality data. Our results show that the excess variance induced by residual gain and bandpass errors dominates over the bias across the range of angular multipoles considered. The amplitude of this variance is comparable to the expected power spectrum upper limit at redshift $z \sim 2.2$, indicating that residual calibration errors can significantly affect the detectability of the 21-cm signal.
We further transform the estimated MAPS into $(k_{\perp}, k_{\parallel})$ space and find the expected foreground wedge structure. Importantly, we observe a significant power leakage into the nominal EoR window, which we attribute to residual calibration systematics. This highlights that even after standard foreground mitigation, time- and frequency-correlated gain errors can introduce mode-mixing and contaminate the EoR window.
Our results emphasise the necessity of improved calibration accuracy and robust characterisation of residual gain statistics for current and upcoming interferometric experiments such as the uGMRT and SKA, to enable reliable detection of the cosmological 21-cm signal.

 }

\keywords {cosmology: dark ages, reionization -- methods: analytical, numerical, statistical -- techniques: interferometric}

\begin{document}
\maketitle
\flushbottom

\section{Introduction}
\label{sec:introduction}

The redshifted 21-cm signal from neutral hydrogen (\HI) is the ideal probe to study large-scale structure of the early universe and its evolution  over cosmic time \cite{1985ApJ...292..371D, 1990MNRAS.247..510S, 1993MNRAS.265..101S, 1995MNRAS.272..544K,  1997ApJ...475..429M}. As the 21-cm signal is weak, directly imaging it at  high redshift is difficult. Alternatively, we use the method of intensity mapping, a  statistical measurement tool that estimates the two-point correlation function or the angular power spectrum \cite{ 2001JApA...22...21B, 2010ARA&A..48..127M} to extract information about the underlying large-scale distribution of \HI \cite{2004ApJ...608..622Z, 2004ApJ...613....1F, 2005ApJ...625..575S, 2006PhR...433..181F}. We can directly measure the visibility function of the sky brightness distribution with radio interferometers, and by correlating visibilities corresponding to nearby baselines and frequencies, we can estimate the multifrequency angular power spectrum of the redshifted 21-cm signal \cite{2001JApA...22..293B, 2004ApJ...615....7M,  2005MNRAS.356.1519B, 2007MNRAS.378..119D}.\\

In the same frequency band as that of the redshifted 21-cm signal, there is radio emission from a variety of other astrophysical sources in  our Galaxy and external galaxies, including compact extragalactic point sources, diffuse Galactic synchrotron emission, and free-free emission. These emissions are  collectively called  foregrounds and are typically several orders of magnitude stronger than the expected 21-cm signal, making its detection  challenging \citep[e.g.][]{1999A&A...345..380S, 2002ApJ...564..576D, 2004MNRAS.355.1053D, 2005ApJ...625..575S, 2008MNRAS.385.2166A, 2008MNRAS.389.1319J, 2012MNRAS.426.3295G, Bernardi}.
Several foreground mitigation techniques have been proposed in the literature to overcome this challenge. One commonly used approach is the {\it foreground subtraction}, where an accurate foreground model is subtracted from the observed visibilities or the reconstructed images, and the 21-cm power spectrum is then estimated from residual data \cite{2006ApJ...648..767M, 2009ApJ...695..183B, 2017NewA...57...94C}.
Another promising approach is the {\it foreground avoidance} technique \citep{2010ApJ...724..526D, 2012ApJ...752..137M, 2012ApJ...745..176V, 2012ApJ...756..165P, 2013ApJ...776....6T, 2014PhRvD..90b3018L, 2014PhRvD..90b3019L, 2016ApJ...818..139T, 2022MNRAS.516.2851P}. Here the  cylindrically averaged power spectrum $P(k_{\perp}, k_{\parallel})$ is estimated, where the contribution from the foreground remains confined to a wedge-shaped region  in the $(k_{\perp}, k_{\parallel})$ plane known as the ``foreground wedge''. The \HI power spectrum  is then  estimated from the rest of the region of the $(k_{\perp}, k_{\parallel})$ plane (referred to as the ``EoR window''). In addition to these, further foreground suppression techniques have been developed to improve the robustness of the measurements \citep{2016MNRAS.463.4093C, 2019MNRAS.483.3910C, 2019MNRAS.483.5694B, 2021MNRAS.507.5310A, 2023MNRAS.520.2094E}.

An important aspect of foreground mitigation is understanding the foreground using observations with various existing radio interferometers.
These observational efforts have quantified different components of the foregrounds, investigated algorithms to mitigate them and  hence placed upper limits on the redshifted 21-cm power spectrum after applying suitable foreground mitigation techniques. Ghosh et al. (2012) \cite{2012MNRAS.426.3295G} use $\sim 10$ hours of observations with the Giant Metrewave Radio Telescope (GMRT)\footnote{See: \url{http://www.gmrt.ncra.tifr.res.in}} at $150$ MHz, where compact foreground sources are subtracted, and report a measurement of diffuse Galactic emission with a strength of $\sim 1000~\rm mK^2$ over the range $k \sim 0.12 - 1.2~h~{\rm Mpc}^{-1}$. Using a deeper GMRT observation of $\sim 40$ hours at the same frequency, Paciga et al. (2013) \cite{2013MNRAS.433..639P} obtain an improved upper limit of $248~\rm mK^2$ at $k \sim 0.5~h~{\rm Mpc}^{-1}$. 
At higher redshifts, Barry et al. (2019) \cite{2019ApJ...884....1B} analyse $\sim 21$ hours of data from the Murchison Widefield Array (MWA)\footnote{See: \url{https://www.mwatelescope.org}} and report an upper limit of $3900~\rm mK^2$ at $k \sim 0.2~h~{\rm Mpc}^{-1}$ for observations corresponding to a redshift of $z \sim 7.1$. Further improvements have been achieved with the Low Frequency Array (LOFAR)\footnote{See: \url{https://www.astron.nl/telescopes/lofar/}}, where Mertens et al. (2020) \cite{2020MNRAS.493.1662M} use $\sim 141$ hours of observations to place an upper limit of $73~\rm mK^2$ at $k \sim 0.075~h~{\rm Mpc}^{-1}$ for the 21-cm signal at $z \sim 9$. 
More recently, Mertens et al. (2025)\cite{2025A&A...698A.186M} report a significant improvement over earlier LOFAR constraints, achieving nearly a two- to fourfold reduction in the upper limits. Their best constraints correspond to $\Delta^2_{21} < (68.7~{\rm mK})^2$ at $k = 0.076~h~{\rm cMpc}^{-1}$ for $z \approx 10.1$, $\Delta^2_{21} < (54.3~{\rm mK})^2$ at the same $k$ for $z \approx 9.1$, and $\Delta^2_{21} < (65.5~{\rm mK})^2$ at $k = 0.083~h~{\rm cMpc}^{-1}$ for $z \approx 8.3$. Sagar et al. (2025) \cite{2025MNRAS.544.3617S} report a recent Band~2 observation towards the ELAIS-N1 field with 32 hours of on-source time, achieve an off-source noise level of $\sim 237$ mJy and measure the power spectrum of the diffuse galactic synchrotron radiation at $120-250$ MHz.
These results highlight the steady progress achieved through improved foreground mitigation and observational strategies, while also emphasising that current measurements are still limited to upper bounds on the 21-cm power spectrum.\\


Foreground contamination and complex instrumental and ionospheric effects pose significant challenges in radio interferometric observations. These effects introduce systematic errors that must be carefully calibrated. Standard calibration procedures, such as primary calibration, self-calibration, and bandpass calibration \citep{1984ARA&A..22...97P, 1992ExA.....2..203W, 2007ITSP...55.4497V, 2009ITSP...57.3512W}, are employed to estimate the complex gains and correct the observed visibilities.
However, the accuracy of calibration is limited by several factors: imperfections in the sky model, limited sensitivity of the instrument,  lack of suitable calibrator sources, antenna gain and bandpass fluctuations, time and space dependence in ionospheric electron density, and radio-frequency interference (RFI), to name a few \citep[e.g.][]{2016MNRAS.461.3135B, 2016MNRAS.463.4317P}. 
A considerable amount of work has been devoted to understanding and characterising these calibration errors, as they can significantly limit the detection of the redshifted 21-cm signal and the astrophysical parameters at reionization \cite{2025JCAP...10..035T}. For instance, studies related to LOFAR-EoR\footnote{Low Frequency Array - Epoch of Reionisation} and  LOFAR-LBA\footnote{Low Frequency Array - long baseline}  experiments have explored the impact of gain errors, polarised foregrounds, ionospheric effects, and calibration-induced biases \citep{2018MNRAS.478.1484G, 2016MNRAS.463.4317P}. Other works have investigated additional effects such as polarisation leakage and ionospheric distortions \citep{2015MNRAS.451.3709A, 2016JApA...37...35A, 2018MNRAS.476.3051A, 2015MNRAS.453..925V, 2016MNRAS.458.3099V, 2016RaSc...51..927M}.
Recent studies have further shown that inaccuracies in sky-based calibration models, as well as departures from perfect redundancy in redundant calibration methods, can introduce gain errors that contaminate the EoR window \citep{2016MNRAS.461.3135B, 2017MNRAS.470.1849E, 2019ApJ...875...70B, 2018AJ....156..285J, 2019MNRAS.487..537O,2022ApJ...941..207K, 2023ApJ...953..136K}. It has also been demonstrated that non-redundancy in the baseline distribution can introduce spectral artifacts that limit  detection of the cosmological signal \citep{2010MNRAS.408.1029L}. Further investigations, particularly for experiments such as Hydrogen Epoch of Reionization Array (HERA) \cite{2017PASP..129d5001D}, show that non-redundancies and unmodelled diffuse emission can bias gain solutions and some may be  addressed through gain-smoothing procedures \citep{2020MNRAS.499.5840D, 2020ApJ...890..122K, 2021MNRAS.506.2066C}. 
Instrumental systematics can also arise from variations in the telescope beam pattern with time and frequency, antenna position errors, and sidelobe effects; all of these can introduce additional distortions in the observed visibilities and affect calibration accuracy \citep{2018AJ....156..285J, 2019MNRAS.487..537O, 2021MNRAS.506.2066C, 2024MNRAS.534.2475C}. These studies collectively highlight that even after foreground mitigation, residual gain and bandpass errors remain a critical limitation, motivating a detailed investigation of their impact on the estimation of the 21-cm power spectrum. 


 Chakraborty et al. (2019b)\cite{2019MNRAS.490..243C} present observations of redshifted 21-cm power spectrum of the European Large Area ISO Survey- North1  (ELAIS-N1) \cite{2000MNRAS.316..749O} field using the upgraded Giant Metrewave Radio Telescope (uGMRT) in the frequency range $300 - 500$ MHz. Using these observations, Chakraborty et al. (2021)\cite{2021ApJ...907L...7C} report upper limits on the 21-cm power spectrum as $(58.87~\mathrm{mK})^2$, $(61.49~\mathrm{mK})^2$, $(60.89~\mathrm{mK})^2$, and $(105.85~\mathrm{mK})^2$ at redshifts $z = 1.96$, $2.19$, $2.62$, and $3.58$, respectively, at $k \sim 1~\mathrm{Mpc}^{-1}$. These measurements place constraints on the product of the neutral hydrogen density parameter and bias, $[\Omega_{\rm HI} b_{\rm HI}]$, yielding values of $0.09$, $0.11$, $0.12$, and $0.24$.
Using the same observation, Pal et al. (2022)\cite{2022MNRAS.516.2851P} employ a foreground avoidance strategy to estimate the power spectrum at $z = 2.28$, and report a $2\sigma$ upper limit of $\Delta^2(k) \leq (133.97~\mathrm{mK})^2$ at $k = 0.347~\mathrm{Mpc}^{-1}$. Further developments in analysis techniques have been explored by Elahi et al. (2023a)\cite{2023MNRAS.520.2094E}. They introduce the Cross-Tapered Gridded Estimator (Cross-TGE) to estimate the multifrequency angular power spectrum (MAPS). In this approach, visibilities from the two parallel-hand polarisation products ($RR$ and $LL$) are correlated to reduce systematics. Applying Cross-TGE to uGMRT Band-3 data at $z = 2.28$, they report a $2\sigma$ upper limit of $\Delta^2(k) \sim (58.67~\mathrm{mK})^2$ at $k = 0.804~\mathrm{Mpc}^{-1}$, corresponding to a constraint of $[\Omega_{\rm HI} b_{\rm HI}]^2 = 7.51 \times 10^{-4} \pm 1.47 \times 10^{-3}$.
In a subsequent analysis, Elahi et al. (2023b)\cite{2023MNRAS.525.3439E} apply foreground removal techniques to the same dataset and obtain a more stringent $2\sigma$ upper limit of $(18.07~\mathrm{mK})^2$ at $k = 0.247~\mathrm{Mpc}^{-1}$, constraining $[\Omega_{\rm HI} b_{\rm HI}] \leq 0.022$.  Elahi et al (2024) \cite{2024MNRAS.529.3372E} further reduce the $2\sigma$ upper limit to $(4.68~\mathrm{mK})^2$ at $k = 0.219~\mathrm{Mpc}^{-1}$ using an improved foreground subtraction technique.


With the advent of new high-dynamic-range radio interferometers, understanding systematic effects and their impact on observations has become increasingly important. In fact, for telescopes such as those of the SKA Observatory systematic effects introduced by calibration errors are among the challenging issues \cite{2016PASA...33...19T, 2019ApJ...875...70B,2025arXiv251025886B,2025arXiv250609533B}. This article is the fifth in the series of our works focusing on the \textit{calibration requirements for Epoch of Reionisation 21-cm signal observations}, where we use existing telescopes like the uGMRT to understand how calibration errors introduce systematic effects into the observed signal, how these effects can affect future observations with the SKA1-Low, particularly for the 21-cm signal and how to mitigate these effects. In our first work \citep[henceforth Paper~I]{2020MNRAS.495.3683K}, we investigated the impact of time-correlated residual gain errors using simulated observations for the GMRT baseline configuration. We demonstrated that such residual gain errors introduce a systematic bias in the estimated power spectrum when using visibility-correlation-based estimators \citep{2014MNRAS.445.4351C}.
 In our second work \citep[henceforth Paper~II]{2022MNRAS.512..186K}, with some simplifying assumptions, we demonstrated that one can analytically estimate the bias and uncertainty in the redshifted 21-cm  power spectrum due to time-correlated residual gain errors and system noise using the Tapered Gridded Estimator (TGE) \cite{2016MNRAS.463.4093C}. These analytical predictions were validated using simulated uGMRT observations, showing excellent agreement between theory and simulations.
In \citep[henceforth Paper~III]{2024JCAP...05..068G}, we verified our assumptions and implemented our analytical framework using uGMRT Band~3 ($300$--$500$ MHz) observations of the ELAIS-N1 field \citep{2019MNRAS.490..243C}. We found that the residual time-dependent gain errors for most antennas follow a Gaussian distribution and exhibit negligible correlation across different Stokes parameters and antennas. Using these empirically derived gain statistics, along with foreground estimates along the same line of sight, we quantified the resulting bias and uncertainty in the detection of the \HI power spectrum. We also provided estimates of the observing time required for a statistically significant detection of the \HI signal under similar instrumental conditions.
In our most recent work \citep[henceforth Paper~IV]{2025JCAP...07..024G}, we extended the analytical framework to incorporate frequency-dependent residual bandpass errors. This allowed us to estimate the bias and uncertainty in the multifrequency angular power spectrum (MAPS)  using the TGE-based MAPS estimator \cite{2019MNRAS.483.5694B}, in the presence of both time- and frequency-dependent calibration errors, particularly in the regime dominated by strong foregrounds. Furthermore, we applied this extended framework to mock observations for SKA1-Low to investigate the combined effects of residual calibration errors on large-scale 21-cm signal detection.


In this work, we thoroughly study the time- and frequency-dependent gain calibration statistics and additional flagging techniques, and implement the analytical framework of Paper~IV using uGMRT Band~3 ($300$--$500$ MHz) observations of the ELAIS-N1 field. We quantify the resulting bias and uncertainty in the detection of the \HI power spectrum in both the $(\ell, \Delta \nu)$ and $(k_{\perp}, k_{\parallel})$ planes and discuss possible reasons for power leakage in the ``EoR window'' in the context of systematic effects.

The paper is organised as follows. In Section~2, we present the analytical framework for estimating the bias and variance of the redshifted 21-cm power spectrum.  In Section~3, we describe the observations and data analysis procedures for the uGMRT Band-3 dataset used in this work. The statistical characteristics of the residual calibration errors are investigated in Section~4. Estimation of the foreground from the same observed data is discussed in Section~5. In Section~6, we present estimates of the bias and variance in MAPS and the cylindrically averaged power spectrum. We present our conclusions in Section~7.

\section{Analytical Estimates of Bias and Variance of the Redshifted 21-cm Multifrequency Angular Power Spectrum}
\label{sec:analyticbiasvar}
In Papers~I, II, III, and IV, we have discussed and demonstrated how calibration inaccuracy introduces errors in visibility that essentially add bias and uncertainty in various estimates of the power spectrum. In this section, we provide a brief discussion on the bias and uncertainty in estimates of  MAPS due to time- and frequency-dependent gain errors. Interested readers can find a detailed discussion and analytical calculation for the same in Paper~IV.

\subsection{Time- and Frequency-Dependent Gain}
Radio-interferometers measure the spatial coherence function, the `visibility', of the astronomical signal modified by the ionosphere and the instrumental response. The modifications can be treated in two terms, one that depends on the direction in the sky and one that is independent of the direction of observation with respect to the centre of the field of view. The direction-independent modifications can be incorporated in a complex factor multiplying the visibility called "gain". In this work, we do not consider the effects of direction-dependent modifications. These will be treated in future work. At a time $t$, at an observation frequency $\nu$, for the $i^{th}$ baseline ${\bf U}_i$, if we consider true sky visibility as $ \tilde{V}^S_i(\vec{U}_i, \nu)$ in the presence of thermal noise $\tilde{N}_i(\vec{U}_i)$, the measured visibility $\tilde{V}(\vec{U}_i, \nu)$ can be given as \citep{1996A&AS..117..137H, SynthesisImaging}
\begin{equation}
  \tilde{V}(\vec{U}_i, \nu) = \tilde{\mathbb{G}}_i(t, \nu) \tilde{V}^S_i(\vec{U}_i, \nu) + \tilde{N}_i(\vec{U}_i),
  \label{eq:measurement}
\end{equation}
where $\tilde{\mathbb{G}}_i(t, \nu)$ is the time- and frequency-dependent baseline-based complex gain. We can safely assume that true sky visibility will not change with time because the observation time is very small in comparison to the time scale of change in the sky signal \citep{1986isra.book.....T}.
A typical approach is to write the time- and frequency-dependent gains separately as 
\begin{equation}
    \tilde{\mathbb{G}}_i(t, \nu) = \tilde{\mathrm{G}}_i(t) \tilde{\mathfrak{B}}_i(\nu),
    \label{eq:gaintnu}
\end{equation}
where $\tilde{\mathrm{G}}_i(t)$ is the time-dependent complex gain  and $\tilde{\mathfrak{B}}_i(\nu)$ is the frequency-dependent gain. The advantages and limitations of this approach are discussed in detail in \cite{2011A&A...527A.106S}. Although alternative methods exist (see, e.g., \cite{2013MNRAS.435..597K}), treating time- and frequency-dependent gains separately is a standard approach and we also adopt it here. For the presently used software-based correlators,   we can assume that the main sources of gains are from modifications by or at  individual antennas  \citep{2017isra.book.....T}, so that we can write 
\begin{equation}
 \tilde{\mathrm{G}}_i(t)  = \langle \tilde{g}_A(t)  \tilde{g}^*_B(t)  \rangle \ ; \ \ \ \ 
    \tilde{\mathfrak{B}}_i(\nu) = \langle \tilde{b}_A(\nu)  \tilde{\sc b}^{*}_B(\nu)  \rangle,
 \label{eq:gain}            
\end{equation}
 where $\tilde{g}_A(t) $, $\tilde{g}_B(t) $ and $\tilde{b}_A(\nu)$, $\tilde{b}_B(\nu)$ are termed the gain and bandpass responses, respectively, of antennas A and B. Here, the gain and bandpass responses are given in terms of the continuous variables time $t$ and frequency $\nu$ and the angular brackets represent the average over integration time and channel width respectively.
We can consider thermal noise $\tilde{N}_i(\vec{U}_i)$ as a Gaussian random variable with zero mean, and there is no correlation across different frequencies channels and different baselines. The standard deviation of the thermal noise per visibility can be expressed  as  \citep{1986isra.book.....T}, 
\begin{equation}
    \sigma_N = \frac{\text{SEFD}}{\sqrt{2\Delta \nu_c\,  T_{\rm int}}}.
    \label{eq:sigma_v}  
\end{equation}
where  $\Delta \nu_c$ is the channel width, $ T_{\rm int}$ is the integration time of visibility measurement and $\text{SEFD}$ denotes the system equivalent flux density. 

\subsection{Residual Gain and Bandpass Errors}
In a typical observation, a visibility is measured at discrete time and frequency steps, owing to finite integration time  and a finite channel width for visibility measurements. The observed data are calibrated for time--dependent gains  by primary calibration followed by  self-calibration and frequency-dependent gain using primary calibrator sources with known frequency response. However, the discretely measured visibilities and the thermal noise present in the measurements limit the achievable calibration accuracy. In the ideal case, once the calibration is applied to the data, the calibrated-visibilities should match the sky-visibilities, rendering all gains as unity in Eq.~(\ref{eq:measurement}). In practice, time- and frequency-dependent gains after calibration can be expressed as (see papers~II and IV),
\begin{equation}
 \tilde{g}(t) = \left [ 1+\delta_{R}(t) + i\delta_{I}(t)\right ] ; \ \ \ \ 
 \tilde{b}(\nu) = \left [ 1+b_{R}(\nu) + ib_{I}(\nu)\right ]
 \label{eq:gaindef}
\end{equation} 
where $\delta $ and $b$ are termed "residual gain error" and "residual bandpass error", $R$ and $I$ denote  the real and imaginary parts. Note that we omit the antenna indices here, although the gains differ between antennas. Further, we assume, (a) the residual gain errors and residual bandpass errors are Gaussian random variables with zero mean and (b) statistics of the gains are similar for all antennas. The assumption of Gaussianity is justified if the residual gains are small and hence the higher order corrections can be neglected. We further assume that there is no correlation between the real and imaginary parts of the residual gain errors of the same or different antennas. We test this assumption comprehensively and examine the variation in the statistics across antennas for a given observation before applying these results. We define the variance and the time- and frequency-correlation function of residual gain errors as, 
\begin{eqnarray}
\sigma^2_{\delta C} &=& \langle \delta^2_{C} \rangle, \ \ \ \ \eta_{C}(\tau) = \langle \delta_{C}(t) \delta_{C}(t+\tau) \rangle / \sigma_{\delta C}^2, \label{eq:gbmod} \\ \nonumber
\sigma^2_{bC} &=& \langle b^2_{C} \rangle, \ \ \ \ \xi(\Delta \nu) = \langle b_{C}(\nu) b_{C}(\nu+\Delta \nu) \rangle / \sigma_{bC}^2,
\end{eqnarray}
where $C $ is the representative of R and I.  \\

\subsection{The Multifrequency Angular Power Spectrum Estimator}
In a typical radio interferometric observation that aims to detect the power spectrum of the \HI emission from the reionisation and post-reionisation eras, the visibilities are recorded for different frequencies channels across a given bandwidth of observation. Datta et al. (2007) \cite{2001JApA...22..293B} show that correlating visibilities across different frequencies and baselines  (also see \cite{2001JApA...22..293B}) allows us to measure the multifrequency angular power spectrum (MAPS) of the sky brightness temperature as a function of both angular multipole ($\ell$) and frequency separation ($\Delta \nu$). The MAPS $C_{\ell}(\Delta \nu) $ is defined as  
\begin{equation}
  C_{\ell}(\Delta \nu) = \langle a_{lm}(\nu) a_{lm}^*(\nu+\Delta \nu) \rangle,
  \label{eq:cldef}
\end{equation}
where $a_{lm}(\nu)$ are the spherical harmonics of the brightness temperature distribution in the sky $\delta T(\hat{n}, \nu)$,  $\Delta \nu$ is the frequency difference between two channels for visibility correlation and angular multipole $\ell$ is analogous to the baseline ($\ell = 2 \pi U$). Though it has been shown that the \HI 21-cm signal from cosmological redshift is not quite ergodic in frequency \cite{Pramanick_2025}, and MAPS needs to be considered as a function of the individual frequencies where the visibilities are correlated, in this work, we assume that the statistical signal in question is ergodic in frequencies and ignore this effect. For a small field of view, the cylindrically averaged power spectrum of the 21-cm signal can be estimated from the MAPS using \cite{2021MNRAS.501.3378P}:
\begin{equation}
  P(k_{\perp}, k_{||}) = r^2 r^{'} \int _{-\infty}^{\infty} {\rm d} ( \Delta \nu) \exp ^{-i k_{||} r^{'} \Delta \nu} C_{\ell} (\Delta \nu),
  \label{eq:cl2pk}
\end{equation}
where $r$ is the comoving distance to the redshift of observation, $r^{'} = {\rm d} r / {\rm d} \nu$, $k_{||}$ is  the component of the wavenumber along the line of sight of observation, whereas, $k_{\perp} = \ell/r$ is the modulus of its component in the sky plane. Bharadwaj et al. (2019) \cite{2019MNRAS.483.5694B} introduced a visibility-based estimator for the MAPS (CLTGE henceforth), that (a) grids the visibilities in the baseline plane and (b) tapers the response of the primary beam to reduce the effect of point sources beyond the tapering window. We present here a brief description of the estimator.
The visibilities are first organised in 2-dimensional discrete grids in the baseline plane for each available frequency channel. The grid spacing $\Delta U$ is chosen to satisfy the condition $\Delta U < (\pi \theta_0)^{-1}$, where $\theta_0$ denotes the angular extent of the telescope field of view in radians. This ensures that an adequate number of baselines are present within a grid and that they remain correlated. During the gridding process, a convolution kernel is applied, which effectively tapers the primary beam response to mitigate the effect of point sources beyond the restricted sky response.
Within each grid, correlations are computed between visibilities corresponding to different pairs of frequency channels
and such correlations are collected in each baseline grid  and frequency separation $\Delta \nu$.  These  estimates are subsequently combined over annular bins of baselines, that is over ranges of angular multipoles $\ell$, to obtain the final estimate of $C_{\ell}(\Delta \nu)$. 
For the case of zero frequency separation, i.e., $\Delta \nu = 0$, the visibility correlations are restricted to pairs belonging to different baselines within the same grid, to reduce the bias from thermal noise. 
We used CLTGE in Paper~IV to demonstrate  the effect of bias and variance arising due to the residual gain- and bandpass errors for the SKA1-Low AA4 configuration. 
In the present work, we use this framework to quantify the bias and excess variance in an archival uGMRT Band~3 observation.

\subsection{Baseline pair fraction}
Every visibility recorded by the radio-interferometer at a time $t$ and frequency $\nu$ is a correlation of the electric fields recorded by a pair of antennas (say  $A$ and $B$) at a baseline vector $\vec{U}$ given by the antenna separation vector projected to the instantaneous sky-plane in units of observing wavelengths. Hence,  a pair of antennas measures visibility  for different baselines at different times. We bring our attention now to the calculation of visibility correlation in a grid in the baseline plane, where many different visibility pairs are correlated. The different types of baseline pairs that can be formed within a given grid for CLTGE are discussed in Paper~IV and listed below. 
\begin{itemize}
    \item {\bf Type 1} Visibility correlations between  the same antenna pair, at different times, and different frequencies, i.e.
      $\langle  \tilde{V}_{AB}(t, \nu) \tilde{V}^*_{AB}(t', \nu')\rangle$. 
     \item {\bf Type 1A} Visibility correlations between the same antenna pair, at the same time, and different frequencies, i.e.
      $\langle  \tilde{V}_{AB}(t, \nu) \tilde{V}^*_{AB}(t, \nu')\rangle$. 
    \item {\bf Type 2} Visibility correlations between antenna pairs having one antenna in common, measured at the same time, and different frequencies, i.e.
    $\langle  \tilde{V}_{AB}(t, \nu) \tilde{V}^*_{AC}(t, \nu')\rangle$. 
     \item {\bf Type 3} Visibility correlations between antenna pairs having one antenna in common, measured at different times, and different frequencies $\langle  \tilde{V}_{AB}(t, \nu) \tilde{V}^*_{AC}(t', \nu')\rangle$. 
    \item {\bf Type 4} Visibility correlations between antenna pairs having no antenna in common, measured at any time.
     $\langle  \tilde{V}_{AB}(t, \nu) \tilde{V}^*_{CD}(t', \nu')\rangle$. 
\end{itemize}
The ratio of a given type of baseline pairs to the total number of baseline pairs in a given grid is defined as the baseline pair fraction. We denote it as $n_{i}$ for different types `$ i $' such as $n_{1}$, $n_{1A}$, $n_{2}$, $n_{3}$ and $n_{4}$. For a given observation, the baseline pair-fractions can be calculated from the observed data and used for subsequent analysis.

\subsection{Bias and Variance in Redshifted 21-cm MAPS}
Assuming perfect calibration and the absence of foregrounds,  Ali et al. (2008) \cite{2008MNRAS.385.2166A} have shown that the visibility-based power spectrum estimators are unbiased and the uncertainty in the power spectrum estimates arises from the sample variance and measurement errors \cite{2008MNRAS.385.2166A}. Any radio frequency emission, other than the 21-cm signal, from the same direction and frequency range of observation is collectively termed the foreground to the 21-cm signal; we denote these with a suffix `$_{FG}$'.


In Papers II and IV,  we introduce an analytical framework to estimate bias and excess uncertainty due to residual gain- and bandpass errors. Here we discuss the final results as presented in those papers for completeness. Bias $\mathcal{B}_{C_{\ell}} (\Delta \nu)$  in the estimates of MAPS can be written as
 \begin{equation}
\mathcal{B}_{C_{\ell}}  =  \left[ (n_{13}\chi_{\rm eff} + n_{12}) \Sigma_2^{\delta+} + (n_{13} + n_{12}) \Sigma_2^{b+} \xi_{\rm eff}\right]  \frac{C_{\ell}}{N_d}
\label{eq:bias}
\end{equation}
where, we have not written the $\ell$ and $\Delta \nu$ dependences explicitly. Here $N_d$ is the number of days of observations. Assuming that observations are carried out for $N_h$ hours each day towards the target field, the total observation time is $T_{obs} = N_h \times N_d$. The quantity $C_{\ell}(\Delta \nu)$ includes both the foreground ($C_{\ell_{FG}}$)  and redshifted 21-cm signal ($C_{\ell_{HI}}$). However, since  ($C_{\ell_{FG}}$) is several orders of magnitude higher \cite{2005ApJ...625..575S}, we can safely use $C_{\ell_{FG}}$ for $C_{\ell}$. Here $n_{13} = 2n_1 + n_3$ and  $n_{12} = 2n_{1A} + n_2 $, $\Sigma_2^{\delta +/-} = \sigma^2_{\delta R} \pm \sigma^2_{\delta I}$ and $\Sigma_2^{b +/-} = \sigma^2_{b R} \pm \sigma^2_{b I}$. Here 
\begin{equation}
\xi_{\rm eff}(\Delta \nu) = \frac{\xi_{\rm R}(\Delta \nu)\, \sigma^2_{b \rm R} + \xi_{\rm I}(\Delta \nu)\, \sigma^2_{b \rm I}}{\sigma^2_{b \rm R} + \sigma^2_{b \rm I}}.
\end{equation}
The variable $\chi_{eff}$ arises due to  the time-correlation in residual gain error and is given as
\begin{equation}
\chi_{\rm eff}(\ell) = \frac{\chi_{\rm R}(\ell)\, \sigma^2_{\delta \rm R} + \chi_{\rm I}(\ell)\, \sigma^2_{\delta \rm I}}{\sigma^2_{\delta \rm R} + \sigma^2_{\delta \rm I}},
\end{equation}
where for `C' as `R: real' or `I:imaginary' and 
\begin{equation}
  \chi_C(\ell) = \frac{1}{T^2_D}\int^{T_D}_{T_{\rm int}} \ \left [ T_D - \tau \right ] \eta_C(\tau) \ {\rm d} \tau.
   \label{eq:chidef} 
\end{equation}
Note,  $T_D = \frac{\Delta U T_{24}}{\ell}$ and $T_{24}$ corresponds to one sidereal day.

 
 The excess uncertainty that arises due to residual gain- and bandpass errors, over the thermal values, can then be written as 
\begin{eqnarray}
 \sigma^2_{E} &=&  2\  \frac{\mathcal{B}_{C_{\ell}}^2}{N_G} + 2 \left [  \Sigma_2^{\delta + } + \Sigma_2^{b+} \right ] \frac{N_2 C_{\ell}}{N_B N_d^2} \label{eq:sigmaE} \\ \nonumber
&+& 4 \left [  (\Sigma_2^{\delta + } + \Sigma_2^{b +} )^2 +  (\Sigma_2^{\delta - } + \Sigma_2^{b -} )^2 \right ]  \frac{C_{\ell}^2}{N_G N_d^2} \\ \nonumber
 &+& \left[ (n_{13}\chi + n_{12}) \Sigma_2^{\delta-} + (n_{13} + n_{12}) \Sigma_2^{b-} \xi \right]^2  \frac{C_{\ell}^2}{N_G N_d^2}, 
\end{eqnarray}
where  $N_G$ and $N_B$ are the number of independent estimates of MAPS and the total number of visibility correlations, respectively, within a baseline annulus in a given day. 
Here the quantity $N_2$ is 
 \begin{equation}
    N_2(\Delta \nu) = \frac{\sigma_{N}^2}{(BW- \Delta \nu)/\Delta \nu_c},
    \label{eq:noisecorr}
\end{equation}
where  $\Delta \nu_c$ is the width of each frequency channel, and BW is the total bandwidth of the observation.

We implement this analytical framework to estimate the bias and excess variance arising in the 21-cm redshifted signal power spectrum of an archival  uGMRT Band~3 observation considering residual gain and bandpass errors. 

\section{Observations and Data Analysis}
\label{sec:observationl}
\begin{table}[h!]
\centering
\begin{tabular}{cc}
\hline
Project code & 32\_120\\
Observation date & May 05, 06, 07, 27 of 2017\\
\hline
Bandwidth & 200 MHz\\
Frequency range & 300--500 MHz\\
Channels & 8192\\
Integration time & 2.68 s\\
Correlations & RR RL LR LL\\
Total on-source time & $\sim 13$ h (ELAIS N1)\\
Working antennas & 28\\
\hline
Flux Calibrator\\
Source & 3C286 (23 Jy), 3C48 (42 Jy)\\
\hline
Phase Calibrator\\
Source & J1549+506 (0.3 Jy)\\
\hline
Target Field\\
Source & ELAIS N1 \\
\hline
\end{tabular}
\caption{Summary of the uGMRT Band-3 observations used in this work.}
\label{tab:observation}
\end{table}
We use  the archival observations of the ELAIS-N1 field using the uGMRT telescope during GTAC cycle 32.  The field was observed over four nights on May 05, 06, 07, and 27 of 2017 with a total on-source time of  13 hours. The  uGMRT  wide-band backend (GWB) was used for a frequency band  of  300--500 MHz with  8192 frequency channels and an integration time of 2.68 s. The relevant observational details are summarised in Table~\ref{tab:observation}.
The raw visibilities are processed following standard procedures, including identification and excision of radio frequency interference (RFI), initial calibration, and imaging. A combined image is generated using data from all observing days and subsequently used as a sky model for direction-independent self-calibration using the {\sc casa} pipeline  \footnote{See: \url{https://casa.nrao.edu}; \citep{2007ASPC..376..127M}}. The final continuum image achieves an angular resolution of $\sim 4.6'' \times 4.3''$ with an rms noise of $\sim 15~\mu$Jy\,beam$^{-1}$.\\ The details of the data analysis, calibration, and imaging are described in \citep{2019MNRAS.490..243C}.

For the present analysis, we estimate antenna-based time-dependent gain solutions from the calibrated visibilities using self-calibration with {\sc solint} set to {\tt int}.  The bandpass response was estimated and the subsequent calibration performed using the standard bandpass calibrator. We do an iterative bandpass calibration  for the bandpass calibrator and preserve  the final bandpass table.  We use these resulting bandpass calibration tables  for further analysis. We estimate the  time-dependent  gains  for each target scan and the frequency-dependent bandpass for all bandpass calibrator scans   for all observing days.


\section{Statistical Characteristics of the Residual Gain Errors}
\label{sec:gainstat}
        

In this section, we  estimate  statistical properties of the residual gain- and  bandpass errors in the data discussed above. We discuss the results using the data from May 07, 2017. 

 \begin{figure}[ht!]
    \centering
    \includegraphics[width=0.47\textwidth]{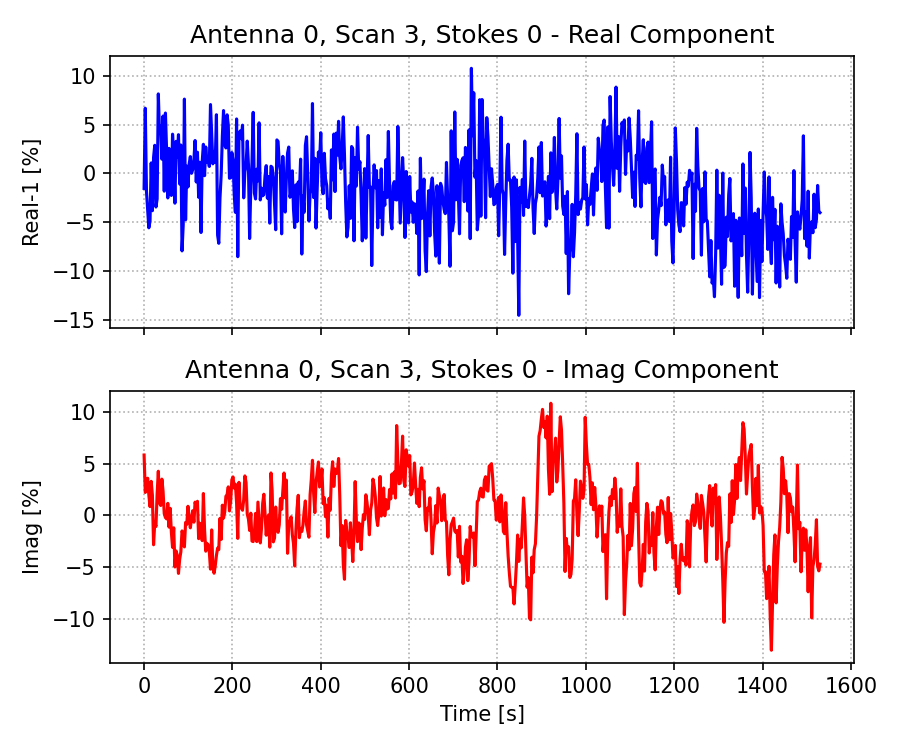}
    \includegraphics[width=0.47\textwidth]{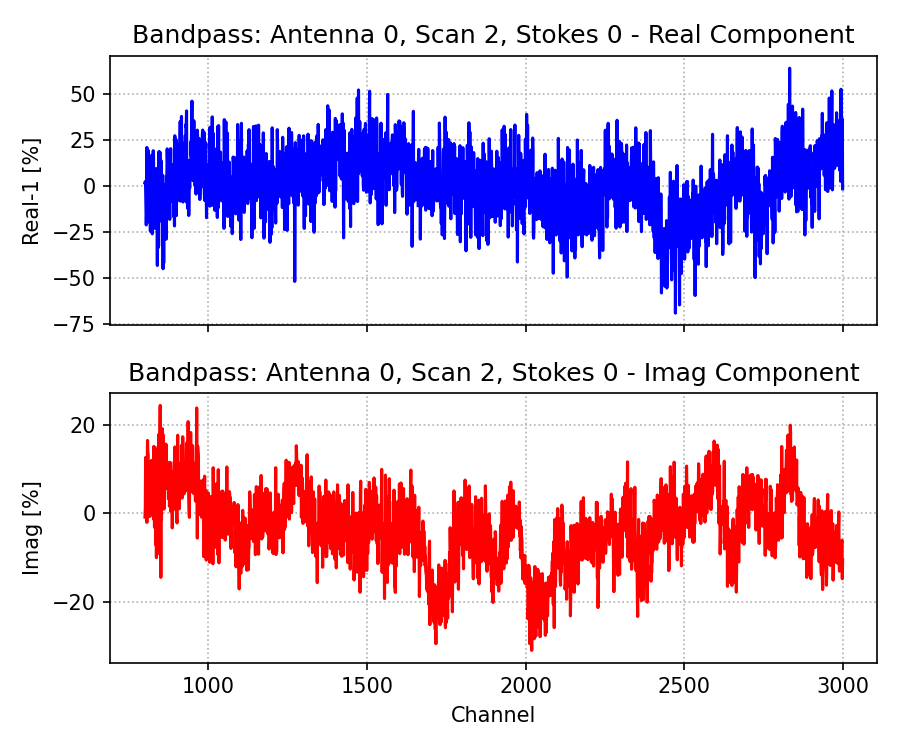}
     \caption{The left panel displays the time series of the residual gain errors (as percentages) for antenna 0 from gain scan 3, for the RR polarisation, showing both real (upper panel) and imaginary (lower panel) components. The right panel shows the corresponding residual bandpass errors for antenna 0 from bandpass scan 2. The observations shown here are from May 07, 2017. Only channels 800--3000 are shown. Note that the real part of the residual gain is ideally unity, whereas the residual error is ideally zero; hence, what is plotted is $100 \times ({\rm real}  - 1)$. }
    \label{fig:gaintime}
\end{figure}

Figure~\ref{fig:gaintime} shows illustrative examples of the residual calibration solutions; the representative statistics used in the subsequent analysis are obtained from target scan~19 and bandpass scan~18. The left panel displays the time series of the residual gain for antenna 0 from gain scan 3, for the RR polarisation, showing both real and imaginary components. The right panel shows the corresponding residual bandpass for antenna 0 from bandpass scan 2. A visual inspection of these plots shows both time and frequency/channel correlations present in the residual gains. Inspecting the bandpass characteristics, we observe that the best usable frequency channels lie between channel number $800$ and $3000$. With a channel width of $\sim 24$ kHz, this gives us about $52.8$ MHz of usable bandwidth, which is adequate for this work.

\subsection{Additional Flagging Based on Gain Statistics}
\label{sec:adgainstat}

For each scan, we first inspect the statistics of gain or bandpass solutions across antennas and time/frequency samples, for both polarisations and the real and imaginary components. The analytical estimates of bias and variance in the MAPS due to residual gain- and bandpass errors assume that (a) the errors are Gaussian random variables, (b) their statistics do not vary significantly across antennas. We investigate the antenna-to-antenna variation of the mean and standard deviation in the time- and frequency-dependent gain errors and  calculate their skewness and kurtosis  for each antenna for a given polarisation and complex component. To check the Gaussianity of the gain- and bandpass errors we also perform  a Kolmogorov–Smirnov (KS) test. Based on these statistics, we further identify certain antenna/time/channel combinations and exclude them from further analysis (flag these combinations). Data points for which either the real or the imaginary parts  exceed a chosen threshold are flagged. Further details of these procedures are given below.

\begin{figure*}[ht!]
    \centering
    \includegraphics[width=0.78\textwidth]{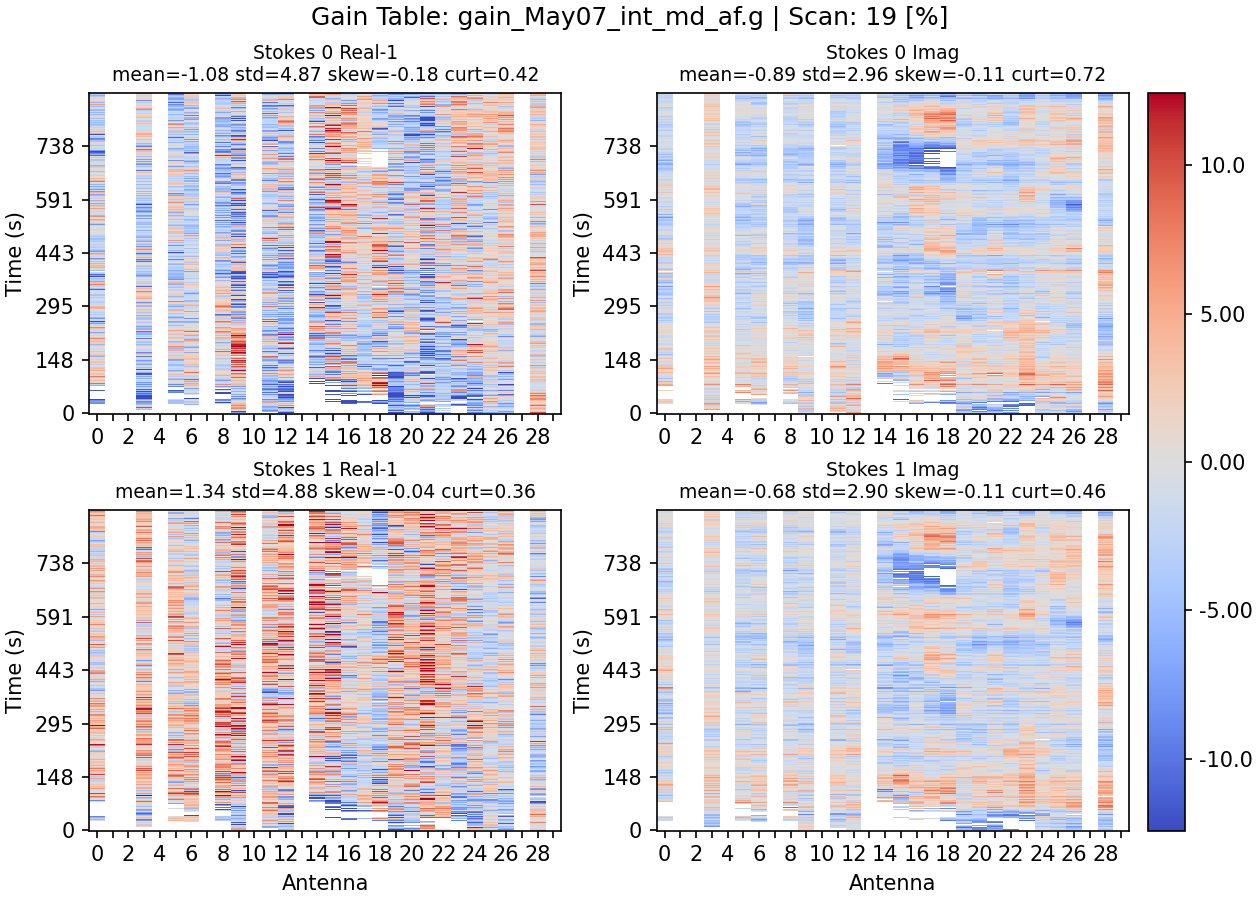}
    \includegraphics[width=0.78\textwidth]{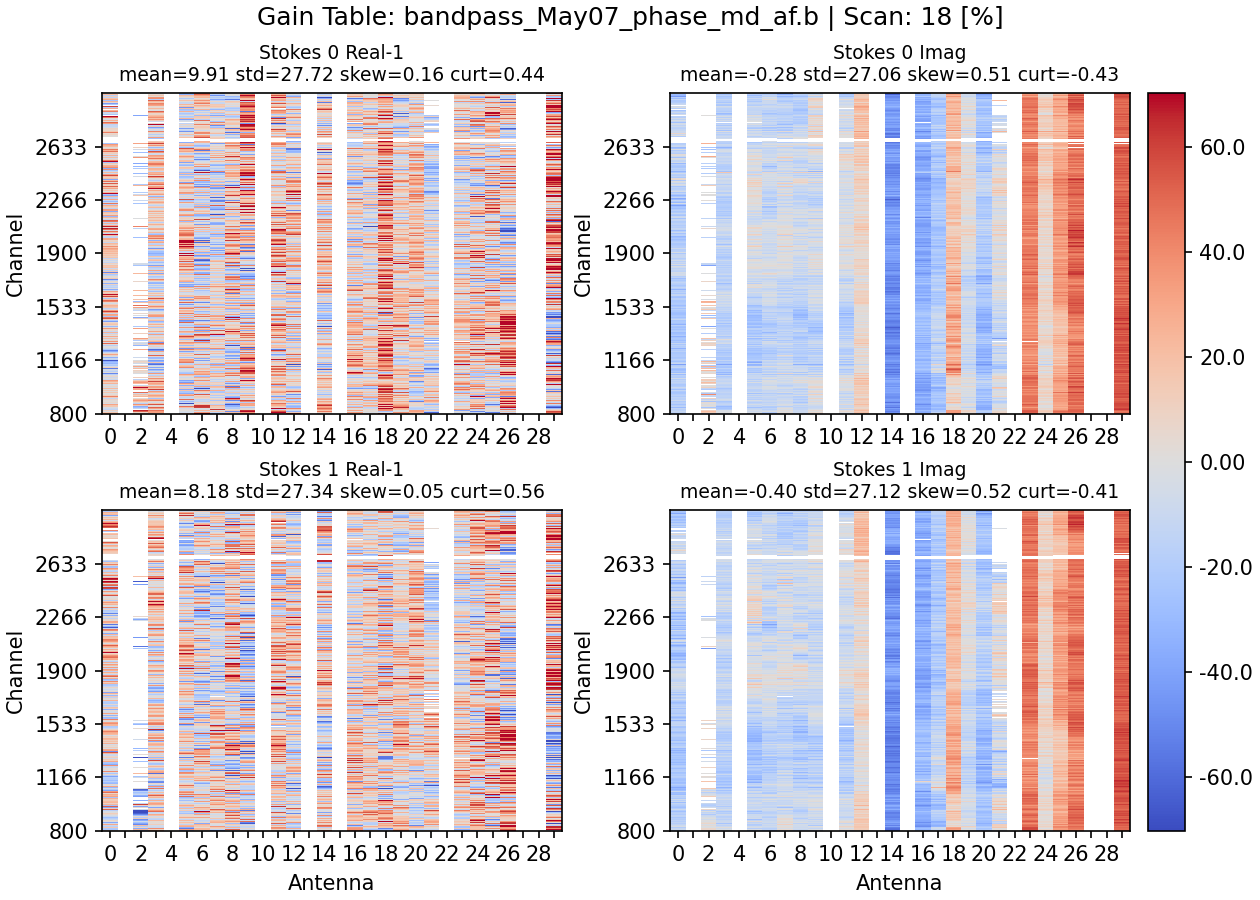}
    \caption{Upper: We plot residual gain errors with colour in the antenna–observation-time plane (as percentages)  for all the antennas and calculate mean, standard deviation, skewness and kurtosis of residual gain errors for RR polarisation real part (upper left), RR polarisation imaginary part (upper right), LL polarisation real part (lower left) and LL polarisation imaginary part (lower right) averaging over all the antennas for scan~19. We then compute the overall standard deviation of the residual time-dependent gain by considering all antennas, both real and imaginary components, and both polarisations.  Lower: Corresponding residual bandpass errors for scan~18.}
    \label{fig:gbcolormap}
\end{figure*}

\begin{enumerate}
    \item For each scan, polarisation and complex component, we compute the overall standard deviation of the residual gain and bandpass by considering all antennas. Data points that exceed a chosen threshold are identified and  flagged. This is iteratively repeated multiple times to progressively remove outliers, choosing different threshold values each time. The process is considered complete when no further points are identified in a given iteration.
    
    \item In the next step, we compute the standard deviation for each antenna separately. Based on this, we apply antenna-specific time- and frequency-domain flagging by removing data points that exceed a predefined threshold. In addition, we flag the entire antenna if a certain percentage of data points were flagged.
    We choose the threshold to be $5-\sigma$ and an antenna is completely flagged if $35\%$ or fewer of its gain or bandpass solutions remain after the previous thresholding.
    
    \item We also investigate the KS-statistics for each scan/polarisation/complex-component combination and for antennas whose KS-statistics value for the residual gain- bandpass-error percentage is higher than $10$, we completely flag those antennas.
\end{enumerate}

The upper plot in Figure~\ref{fig:gbcolormap} shows the time-dependent gain solutions in percentage using colormap after the above procedure. The vertical axis denotes time, while the horizontal axis gives the antenna number. The left and right columns give the real and imaginary components, whereas the upper and lower rows show the RR and LL polarisations ( given as software indices 0 and 1 in the figure titles) respectively.  Summary statistics are shown above each panel. A corresponding figure for the bandpass solutions is given in  the lower plot of the same figure, where the vertical axis gives the frequency channel of observation. The flagged regions in the figures are in white, the vertical white bands show the antennas that are completely flagged.

\begin{figure*}[ht!]
    \centering
    \includegraphics[width=0.9\textwidth]{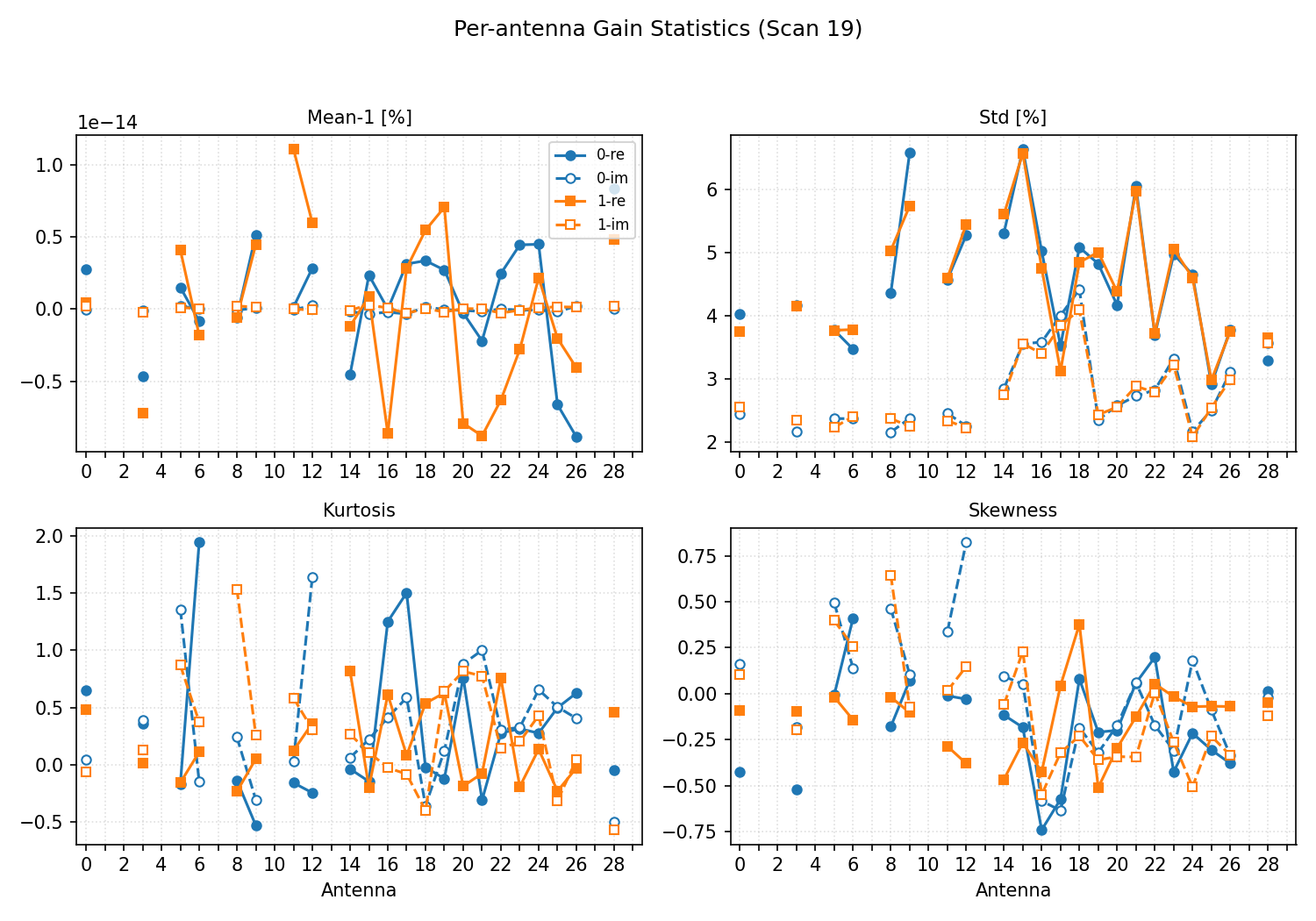}
    \caption{Mean (upper left), standard deviation (upper right), skewness 
    (lower right) and kurtosis (lower left)  of the time-dependent residual gain errors for the two polarisations and complex components are plotted for different antennas. }
    \label{fig:gainsummary}
\end{figure*}
\begin{figure*}[h!]
    \centering
    \includegraphics[width=0.9\textwidth]{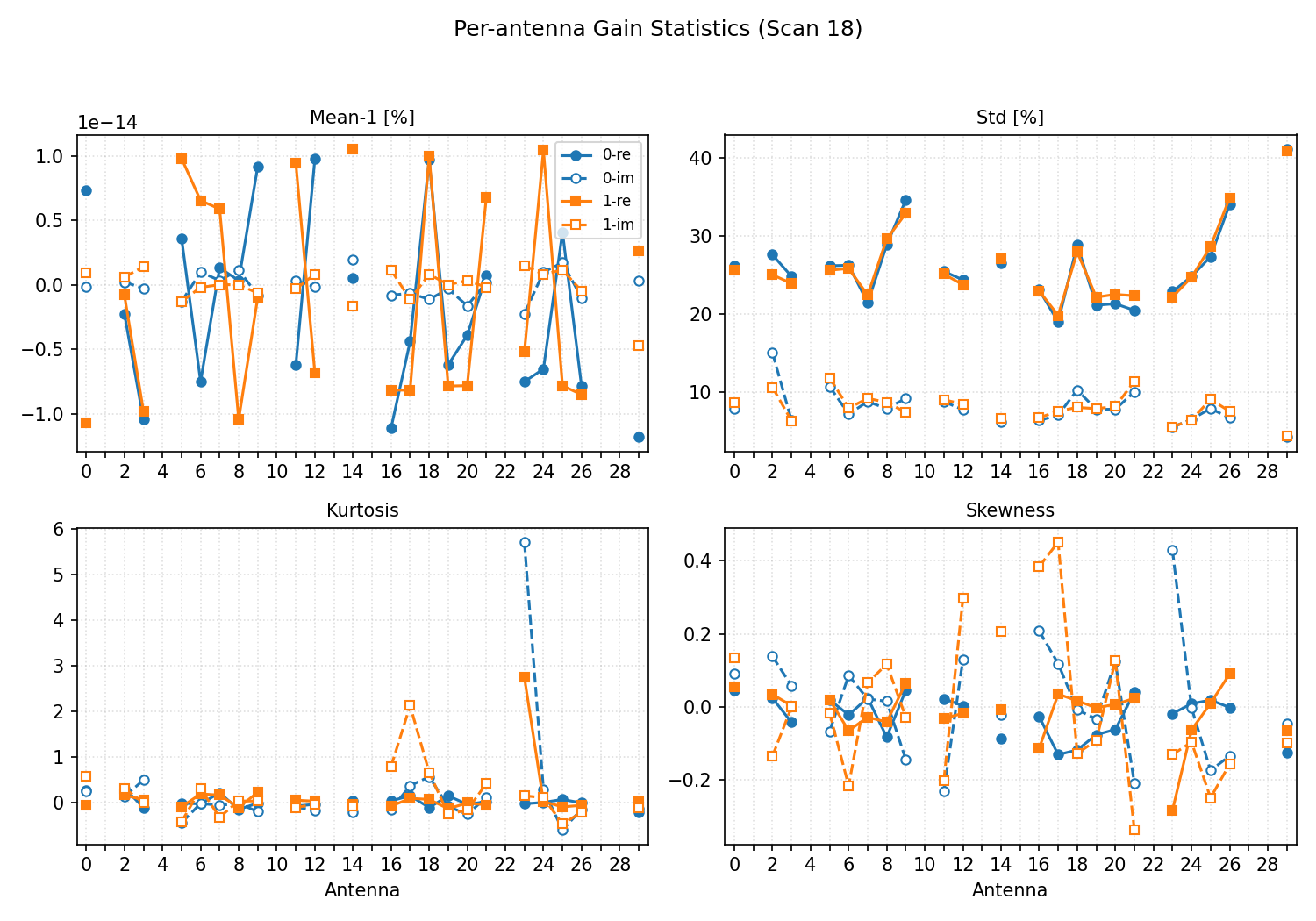}
    \caption{Same as Figure~\ref{fig:gainsummary} for residual bandpass errors.}
    \label{fig:bpasssummary}
\end{figure*}

We observe that there is only about $3-5\%$ variation across different antennas/time for the residual gain errors, whereas for the residual bandpass errors the variation is as large as $\sim 25\%$. Note that the time-dependent gain-calibration is done with the target source with iterative self-calibration, whereas the bandpass calibration is done only with respect to the calibrator source. The higher value of the residual bandpass error suggests a need for a better bandpass calibration strategy. We also observe the existence of correlated residual gain in both time- and frequency-dependent calibration errors. It is particularly interesting that for the imaginary part of the time-dependent calibration error there is visible antenna-to-antenna correlation for the larger antenna numbers (the arm antennas of the uGMRT).

The first four moments (mean, standard deviation, skewness and kurtosis) of the residual time- and frequency-dependent calibration errors for both polarisations and complex components are plotted for different antennas in Figures~\ref{fig:gainsummary} and \ref{fig:bpasssummary}. Here, we have divided the antenna gain and bandpass mean values by the mean value of gains across all antennas to bring out the antenna-to-antenna difference. The statistics are calculated for $\%$ gain errors for the mean and standard deviation and the skewness and kurtosis are weighted by the standard deviation and are dimensionless. For the time-dependent gain errors we observe that the percentage variation of the real part of the mean and standard-deviation are significantly higher than that of the imaginary part, however, neither the real nor the imaginary part exceeds $\sim 5\%$.  Neither the skewness nor the kurtosis values are significant  for any of the antennas. This validates our assumption  that the time-dependent gain errors here (a) are approximately Gaussian and (b) have statistics that do not vary significantly across antennas. The real parts of the residual bandpass errors have a relatively large standard deviation (about $25\%$) for all the antennas, whereas the  standard deviation for the imaginary part is about $10\%$; however, here too, we do not see much antenna-to-antenna variation. The skewness and kurtosis values are mostly insignificant (maximum kurtosis of $\sim 5$) for the antennas,  suggesting Gaussianity within the bandpass errors. 

Furthermore, we note that, in all cases above, the statistics of the residual gain- and bandpass errors are qualitatively similar for the RR and LL polarisations.

\begin{figure}[h!]
    \centering
    \includegraphics[width=0.47\textwidth]{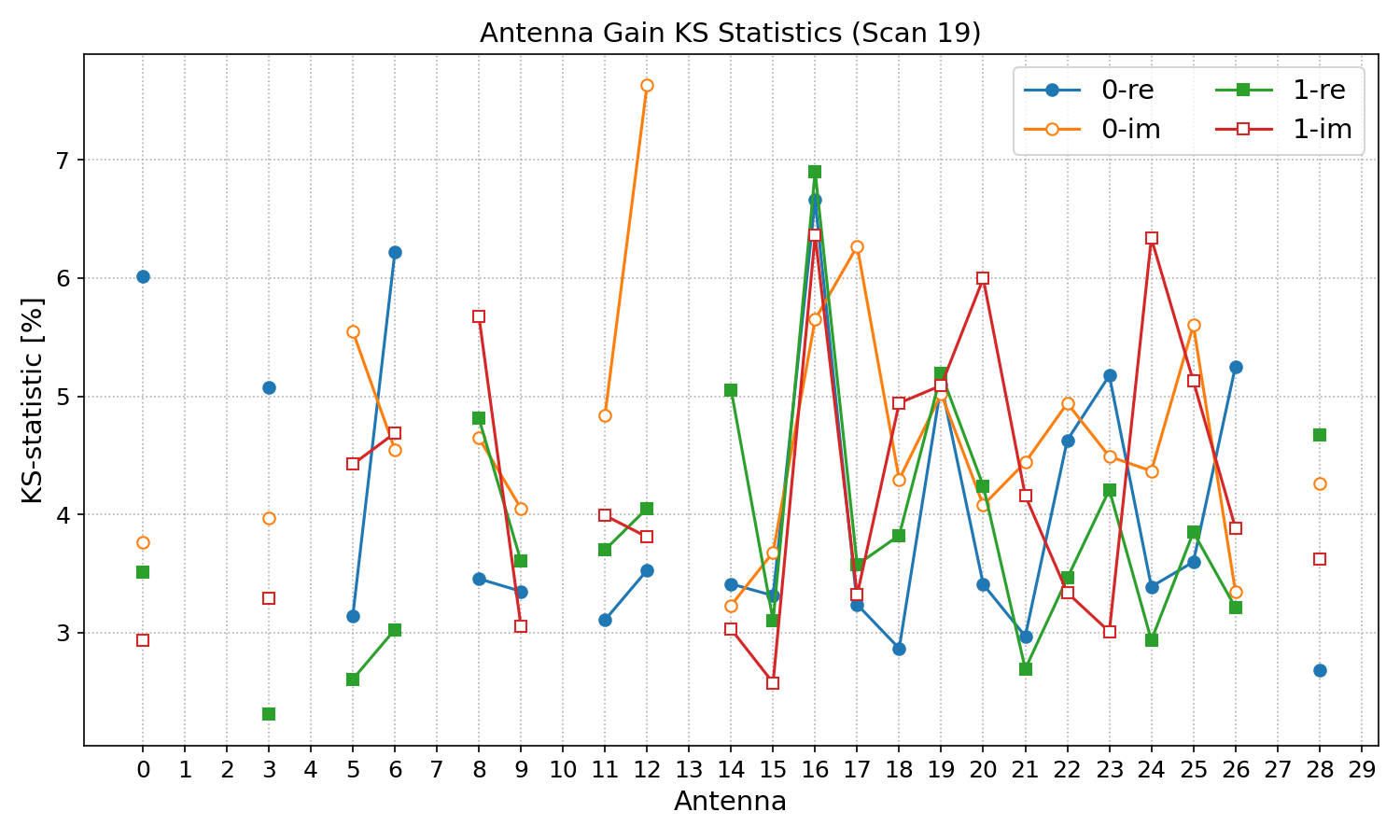}
    \includegraphics[width=0.47\textwidth]{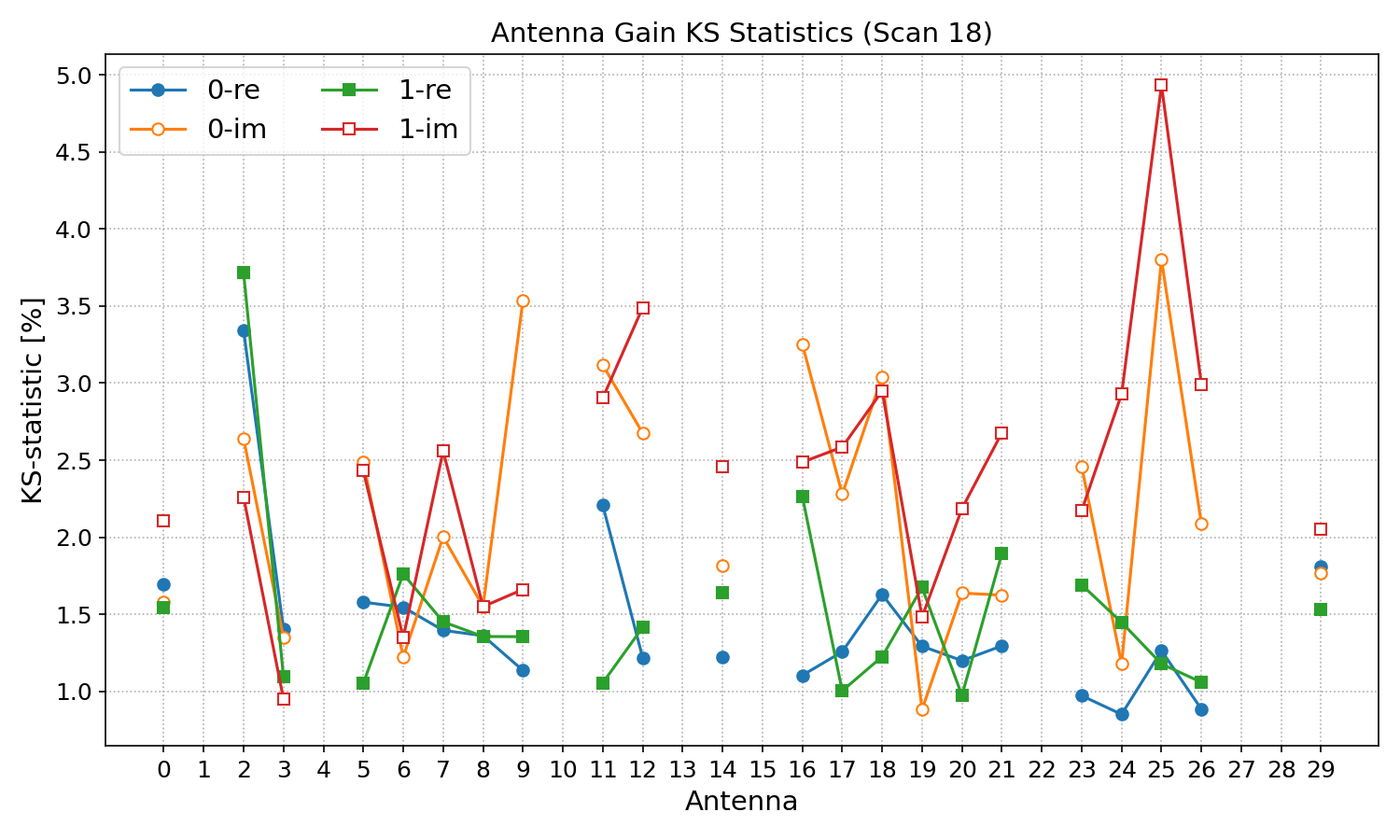}
     \caption{KS statistics values are plotted against antennas. The quoted percentage values of KS statistics correspond to the probability of deviation from a Gaussian distribution. For all the antennas, polarisation, real or imaginary parts of residual calibration errors, it is less than $~7\%$.}
    \label{fig:ks}
\end{figure}

To further quantify the statistical nature of the residuals, we perform a Kolmogorov-Smirnov (KS) test on both the gain and bandpass residual errors. The results are presented in the left and right panels of Figure~\ref{fig:ks}, respectively. The KS statistics indicate that the residual gain distributions are consistent with Gaussianity. The quoted percentage values correspond to the probability of deviation from a Gaussian distribution, and are found to be sufficiently small to justify the Gaussian assumption adopted in our analytical framework.

In the scope of the present work, we use a chosen target source scan and its nearby bandpass scan as representative scans to use the gain and bandpass statistics. We investigated the statistics of residual gain errors for all four days of observations for all available scans. We note that, for observations on May 05, only three target scans were present and we do not use these for our analysis. The summary statistics of the residual gain errors for the target scans of May 06, 07 and 27 are shown in Figures~\ref{fig:gsumm06}, \ref{fig:gsumm07} and \ref{fig:gsumm27} of the Appendix. Note that for these three days there are a total of 8, 8 and 9 target scans respectively. For each scan we show the mean of the standard deviation of the residual gain errors across all antennas with filled blue circles; the error bars represent the antenna-to-antenna variation of the standard deviation. We also show the summary for the skewness (green square) and kurtosis (red open circle) to check the overall Gaussianity of the given scan. Four rows represent the combinations of the two different polarisations and the two complex components as mentioned, where software index $S0$ denotes the RR polarisation and $S1$ denotes the LL polarisation. We found that the best gain solutions are from 9th scan of May 06, 19th scan of May 07 and 10th scan of May 27. We choose the 19th scan of May 07 as the representative scan for this analysis. Corresponding to this scan, we choose the 18th scan of the bandpass calibrator from May 07 as a representative scan for the residual bandpass errors. Note that all the statistics for the residual gain and bandpass errors are demonstrated for these scans in the present section.

\subsection{Time and Frequency Correlation}
To investigate the time- and frequency-correlation in the residual gain- and bandpass errors we use the normalised structure function of order two defined for two random fields $f(x)$ and $g(x)$ having   standard deviations of ($  \sigma_f , \sigma_g $) as
\begin{equation}
S_2(r) = \frac{\langle [f(x+r) - g(x) ]^2 \rangle}{2 \sigma_f \sigma_g}.
\label{eq:s2def}
\end{equation}
If the fields $f$ and $g$ are the same, the structure function gives an estimate of the autocorrelation of that field; otherwise, it gives an estimate of the cross-correlation. 
For the normalised autocorrelation case, the structure function is independent of the mean and is related to the two-point correlation function defined in Eq.~(\ref{eq:gbmod}) such that their sum is unity. Thus, a positively correlated random field has a normalised autocorrelation structure function below unity. Here, both time-correlated residual gain errors and frequency-correlated residual bandpass errors are expected to have the structure function less than unity for lower time lag or frequency separation. We expect the structure function to approach unity at larger time lags or frequency separations. We use the structure function to quantify the correlation in residual gain- and bandpass- errors first for a given antenna/polarisation/complex component (auto-correlation) and then across different antennas (cross-correlation). The structure functions are shown as a function of time lag $\tau$ for residual gain and as a function of channel separation for residual bandpass errors. Note that each channel separation corresponds to $24$ kHz in our observation.

\subsubsection{Auto-correlation}
\label{subsubsec:auto}
\begin{figure}[ht!]
    \centering
    \includegraphics[width=0.95\textwidth]{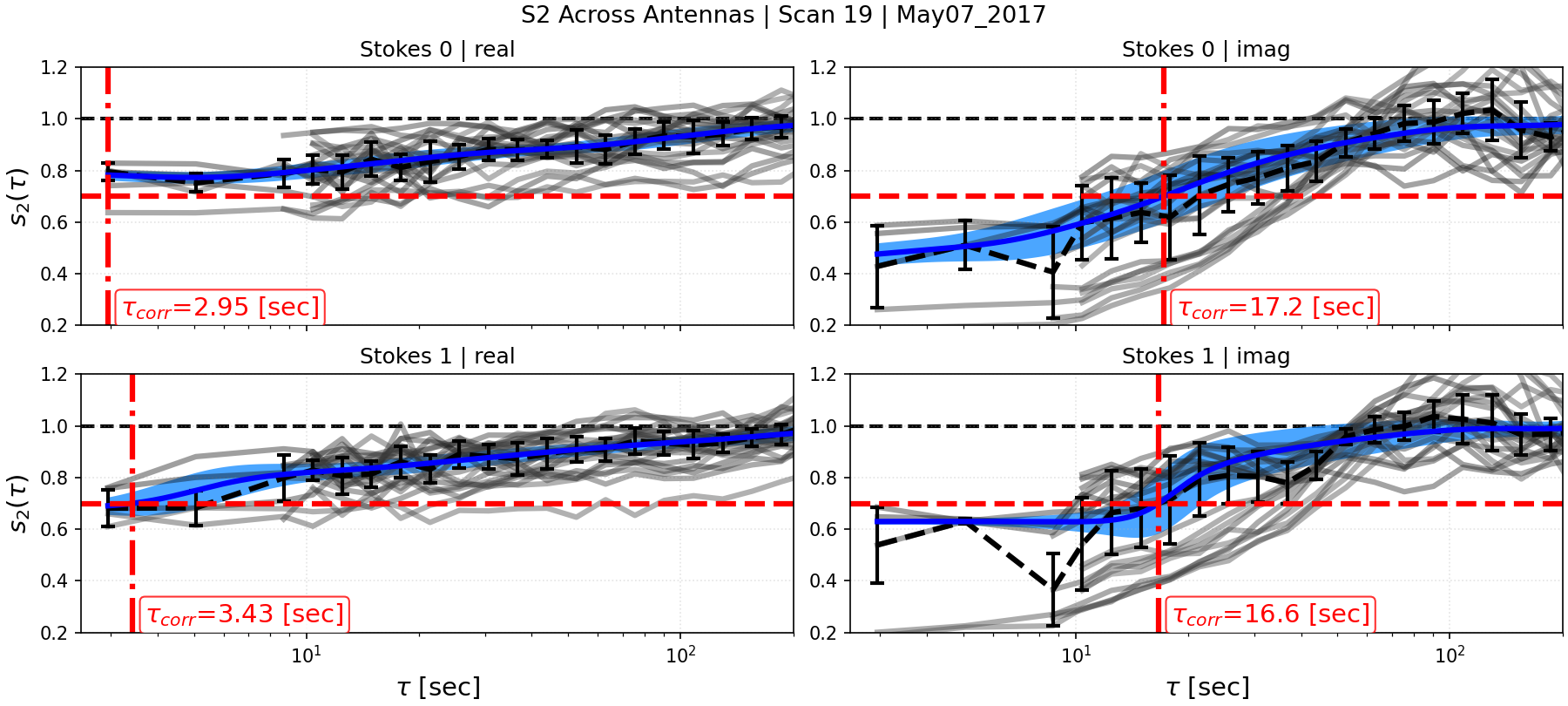}
     \caption{Left and right columns show the structure functions for the real and imaginary values of residual gain errors. The top row is for the RR polarisation (0) and bottom row for the LL polarisation (1). We plot the structure function for individual antennas in grey as a function of lag $\tau$. Median values across different antennas and median absolute deviation values are plotted as a black dashed line  with error bars. The blue line and the blue shaded regions show the Gaussian process regression mean and correlation values. The horizontal red-dashed line corresponds to a threshold of $0.7$; we draw a vertical line denoting the correlation time $\tau_{corr}$ and the corresponding value is written in the margin.}   
      \label{fig:tcorrfun}
\end{figure}
\begin{figure}[ht!]
    \centering
    \includegraphics[width=0.95\textwidth]{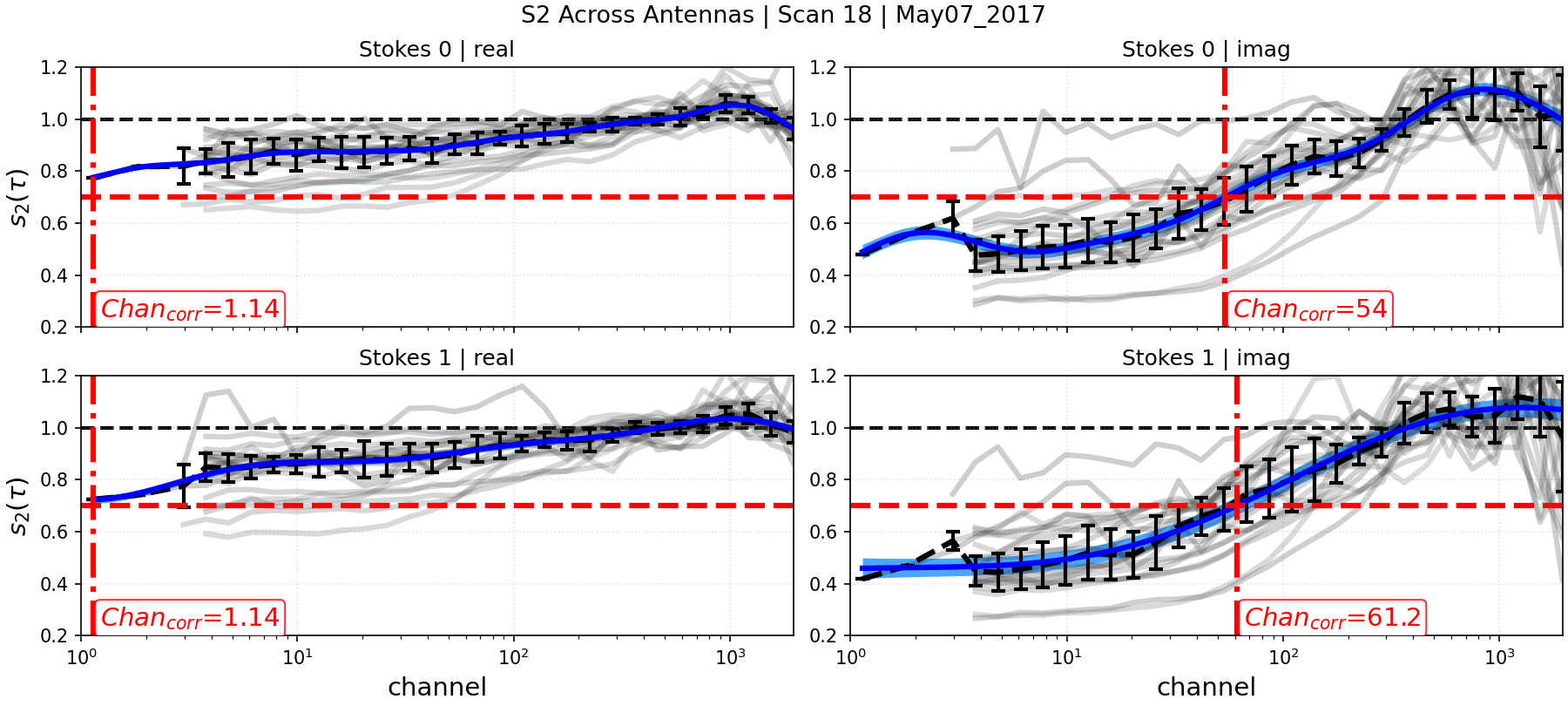}
     \caption{Here we study autocorrelation in residual bandpass errors and calculate how many channels are correlated. The methodology and plot description are the same as in Figure~\ref{fig:tcorrfun}.}
    \label{fig:bpcorrfun}
\end{figure}

For each polarisation and complex component of the residual gain errors we estimate the structure function for all unflagged antennas. These are shown with grey solid lines in  Figure~\ref{fig:tcorrfun}. Note that the analytical estimates we use here assume that the statistics of the residual gain errors are similar across all antennas. We estimate the median values of the structure function and the median absolute deviation across all unflagged antennas. These are shown using a black dashed line and error bars. Note that in most of the cases the variation across the antenna is not significant and within $10\%$. Next we use a Gaussian Process Regression (GPR) model with a Gaussian kernel to find GPR mean and correlation fit (shown with a solid blue line and shaded regions). We use the GPR mean model as a representative structure function for the corresponding polarisation and complex component. Figure~\ref{fig:bpcorrfun} shows the structure functions for the residual bandpass errors as a function of the channel separation. We use the same approach to investigate the autocorrelation in residual bandpass errors.

We define the correlation time or correlation channel separation for the residual gain- and bandpass errors     as the time/channel separation  for which the structure function is less than a threshold value of $0.7$. The horizontal red dashed line in the plots shows this threshold value and the vertical red dash-dotted line indicates the correlation time or channel separation.  If the structure function is found to have a value  $\ge$ threshold in the first bin itself, then the edge of the first bin is considered as the correlation time or channel separation. Note that for the auto-correlations, we have not used the correlation time/channel separation  values as estimated here. We observe that for the real part the correlation times as well as correlation channel separations are significantly smaller and stay within two integration times of observation or channel width, whereas for the imaginary part there is a significant residual  correlation. In all cases we see that there is no significant difference in the gain- and bandpass-statistics between the RR and LL polarisation components.

\subsubsection{Cross-Correlation}
\label{subsubsec:cross}
\begin{figure}[ht!]
    \centering
    \includegraphics[width=0.85\textwidth]{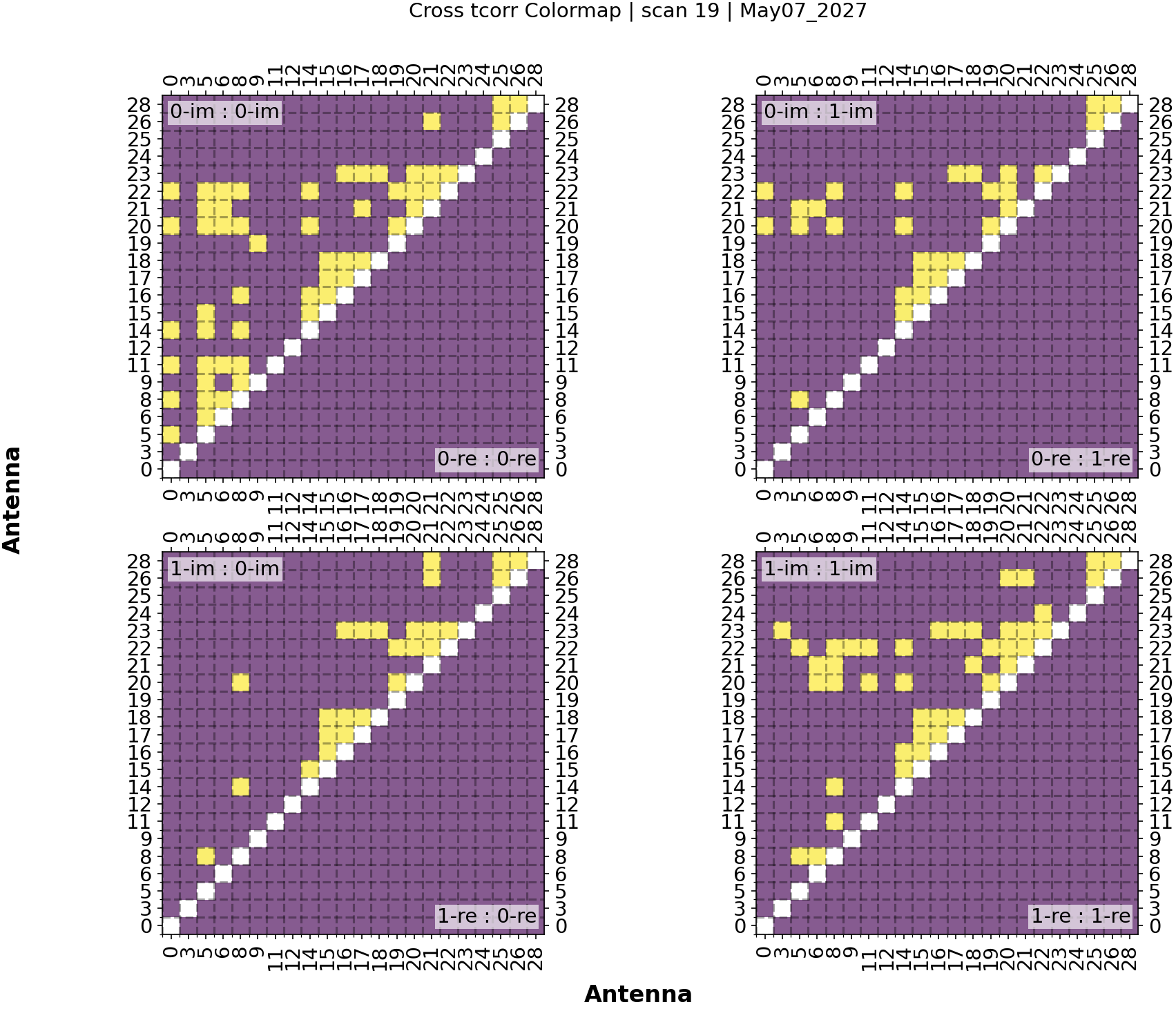}
     \caption{In each of the four panels, the upper-left triangles show the correlation-time  of imaginary-imaginary components, whereas the lower-right triangles show the correlation-time of real-real components for the residual gain errors. Cases with no significant correlations are shown in purple and the yellow boxes represent  correlation time greater than twice the integration time of observation. The four panels correspond to different polarisation combinations (see labels 0,1) for the correlation. The antenna numbers are marked in the margin of each panel.}    
     \label{fig:tcrosscorr}
\end{figure}
\begin{figure}[ht!]
    \centering
    \includegraphics[width=0.85\textwidth]{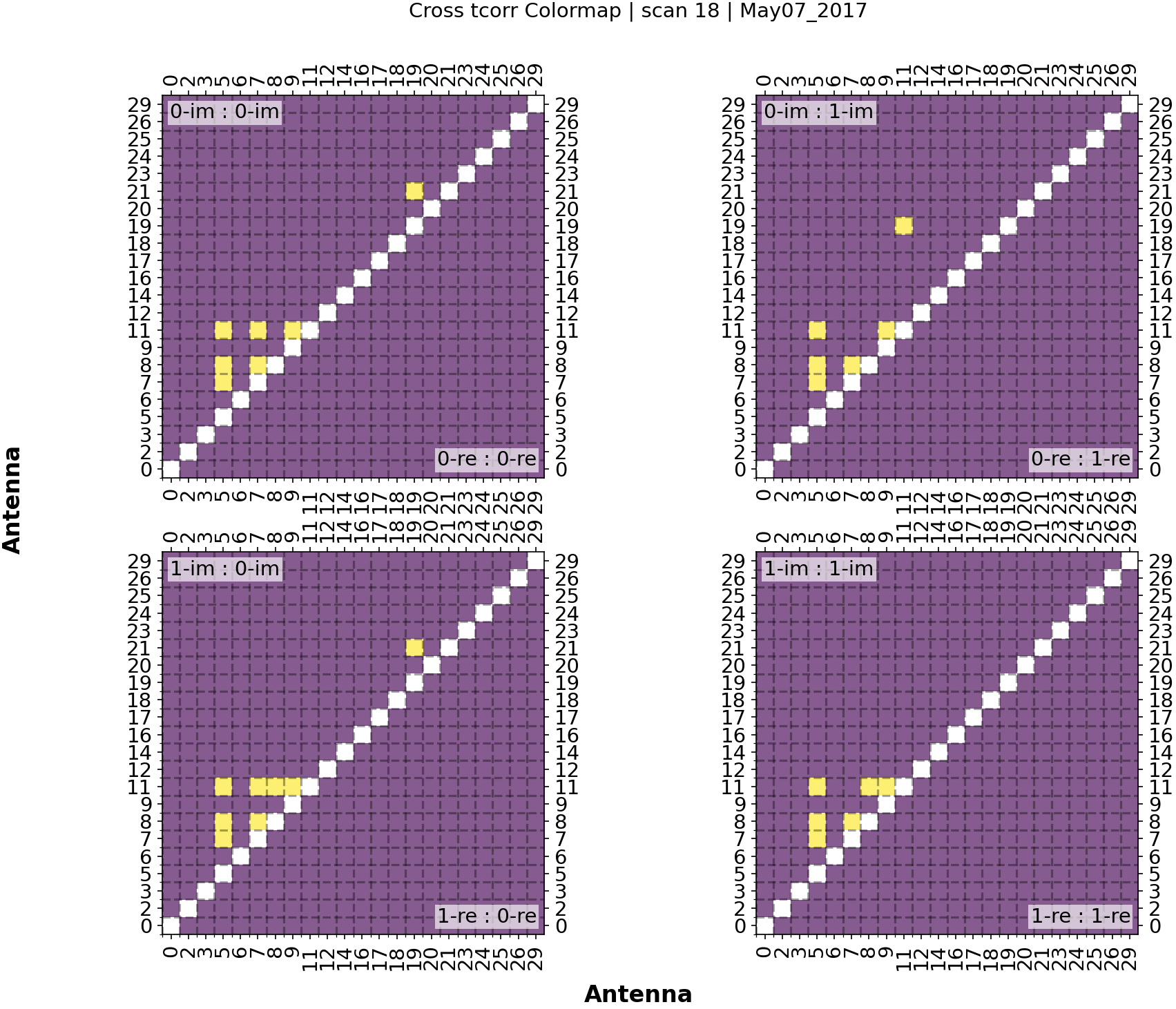}
     \caption{Same as Figure~\ref{fig:tcrosscorr} for residual bandpass errors.}
    \label{fig:bcrosscorr}
\end{figure}

In this work, we assumed that the residual gain- and bandpass errors are Gaussian random variables, and hence only two-point correlations between the gain and bandpass of any two antennas need to be estimated. Let us first discuss the case for residual gain. The effect of cross correlation between the bandpass of two antennas is the same as that of gain. While estimating the cross correlation between the residual gain of the two antennas, we need to consider that the residual gain components of a given antenna can have either of the RR or LL polarisation and either of the real or imaginary of the complex components. This gives us  four functions  for the residual gain error of a given antenna. Hence, for an antenna pair, there exist 16 different combinations of the residual gain cross correlations. We found that the correlation time is consistent with the integration time while the gains from the two antennas are of different complex components, i.e, there is no measurable cross correlation between them. We estimate the cross-correlation time ( as discussed in previous sub-section) for all the other pairs and note the pairs that have significantly larger correlation time than the integration time of the observation. Figure~\ref{fig:tcrosscorr} shows such cases of antenna/polarisation/complex component combinations in yellow. In each of the four blocks, two possible correlation pairs are shown in two triangular patterns. The upper-left triangles show correlations for imaginary-imaginary pairs for all four polarisation combinations for the four panels, whereas the bottom-right triangles show for the real-real pairs. In each triangle , every block corresponds to an antenna-pair marked in the margin of the block. The diagonal lines are kept white as they show self correlation. 
We further observe that none of the real–real pairs show any significant correlation. There are some pairs showing cross-correlations, particularly from the arm antenna of the uGMRT, however, the correlation times were rather short compared to what can affect the analytical calculation significantly. We neglect these effects for further calculation. This is an interesting observation and needs to be investigated in future work to improve calibration algorithms for uGMRT.

Figure~\ref{fig:bcrosscorr} presents the corresponding behaviour for the residual bandpass errors. Similar to the  gain errors, there is no significant cross correlation when  the bandpass from the two antennas are of different complex components or from the real  components alone. We find that most cross-correlations, in general, remain weak, consistent with the assumptions made for the analytical estimates of the bias and uncertainty of the CLTGE estimates of the MAPS.

\section{Foreground estimation}
\label{sec:fg}
\begin{figure*}[ht!]
    \centering
    \includegraphics[width=0.95\textwidth]{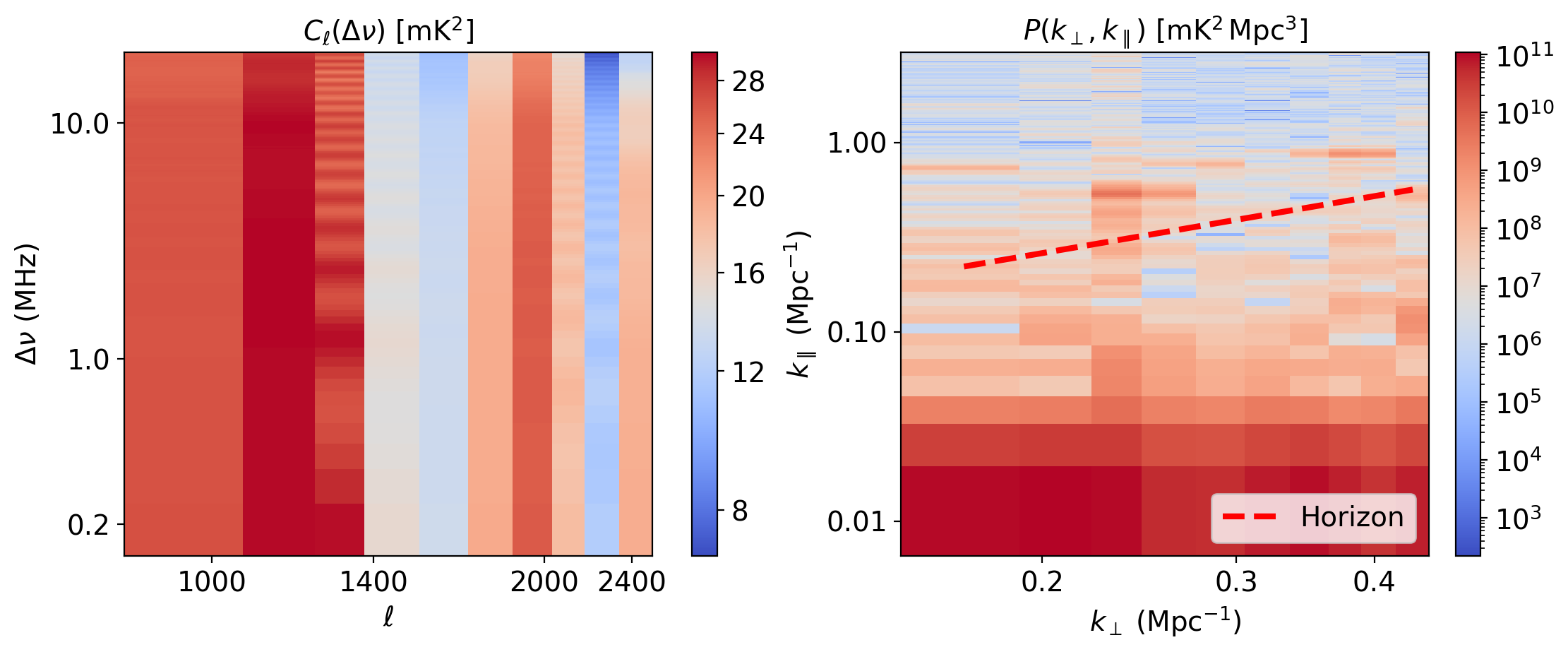}
    \caption{The estimated MAPS  (left) and the cylindrically averaged power spectrum (right) of the foreground.}
    \label{fig:fg}
\end{figure*}

Alongside the residual gain- and bandpass error-statistics from the observed data, to estimate the bias and variance in the MAPS, we use an estimate of the foreground along the same direction of observation. 
We use estimates of the  foreground  MAPS  obtained from uGMRT observations towards ELAIS-N1 carried out on the 3rd of May 2024 (observation id: $46\_032$).  Here we give a brief description of the data reduction and foreground estimation procedures. A detailed account of the analysis is currently being prepared and will be reported separately (Tripathy et al., in preparation). The data are first cleaned of RFI using  Median Absolute Deviation (MAD) statistics  and then passed through an automated direction-independent calibration and flagging pipeline
that follows an improved  framework from  \cite{2019MNRAS.490..243C}. 
 We further use the Cross-Tapered Gridded Estimator (Cross-TGE henceforth)  \cite{2023MNRAS.520.2094E} to  estimate the MAPS $C_\ell(\Delta\nu)$ in the direction of ELAIS-N1 and subsequently the cylindrical power spectrum $ P(k_\perp, k_\parallel)$.  Note that originally,  Cross-TGE tapers the response of the primary beam and hence reduces the effect of the bright sources near the edge of the primary beam and beyond that are not well modeled while imaging. However, the residual gains are corrupted with the full interferometric response. Hence, here we refrain from tapering the antenna response and  we use a Gaussian window function with a relatively large full width at half maximum (FWHM) of $120 \, {\rm arcmin}$,  where the FWHM of the uGMRT primary beam $\approx 62 \, {\rm arcmin}$ at 444 MHz. We have selected a 25~MHz bandwidth of data around 444~MHz, with a frequency resolution of 97.6~kHz. The baseline range is restricted to $ 143\lambda \leq U \leq 398\lambda$ corresponding to angular multipoles of $898 \leq \ell \leq 2501$, which is divided into 10 equally spaced linear bins. This probes the $k_\perp = \ell/r$ range of $ 0.169 \leq k_\perp \leq 0.423 \, \rm{Mpc}^{-1}$  and  $0 < k_\parallel < 3.32 \,\rm{Mpc}^{-1}$, with a resolution of $\Delta k_\parallel = 0.013 \, \rm{Mpc}^{-1}$ for estimated $P(k_\perp, k_\parallel)$.
 
 The left panel of Figure~\ref{fig:fg} shows the estimates of the multifrequency angular power spectrum of the foreground $C_{\ell}(\Delta \nu)$ as a function of the angular multipoles $\ell$ and frequency separation $\Delta \nu$. Since we expect  the MAPS to be correlated to higher extent in $\Delta \nu$ than what is measured here, we use a Blackman-Harris window along the $\Delta \nu$ axis  and use eqn~(\ref{eq:cl2pk}) to find estimates of the power spectrum. The window is normalized to unity at $\Delta \nu =0$. The right panel shows the estimates of the cylindrically averaged foreground power spectrum $P(k_{\perp}, k_{||})$ as a function of the $k$ modes.  We see that the values of the estimated $C_\ell(\Delta\nu)$ vary in the range $3.7-31.0~{\rm mK}^2$ across the $(\ell, \Delta\nu)$ space. We see a nearly constant amplitude of $\sim 8 - 10\,{\rm mK^2}$ on top of which there is an oscillatory pattern. This behaviour is typical of Galactic diffuse emission, as well as extragalactic point source foregrounds present in the data. The point source foregrounds, which are away from the phase centre of the observation, typically show the oscillatory pattern (see, e.g., \cite{2022MNRAS.516.2851P} for a detailed discussion). The values of $\mid P(k_\perp, k_\parallel) \mid$ are in the range $\sim  10^2  -  10^{11} {~\rm mK^2\, Mpc^3}$. We see that a significant part of the power is localised inside the theoretically predicted foreground wedge boundary (dashed line), and there seems to be a relatively cleaner window for 21-cm estimation. These results are in line with previous measurements of the estimated power spectrum using uGMRT Band 3 data \citep{2021ApJ...907L...7C, 2022MNRAS.516.2851P, 2024MNRAS.529.3372E}. 
 
We note that the analytical expressions given here for the bias and variance of MAPS do not consider any uncertainty in the estimate of the foreground itself. We acknowledge that the uncertainty in foreground estimates is expected to modify the systematics in MAPS estimates. However, the proportional errors from the foreground estimates are expected to be smaller than those associated with the gains; at present, we neglect the uncertainty in the foreground estimates.

\section{Results and Discussion}
\label{sec:result}

\subsection{Bias and uncertainty in MAPS}
\begin{figure}[ht!]
    \centering
    \includegraphics[width=0.95\textwidth]{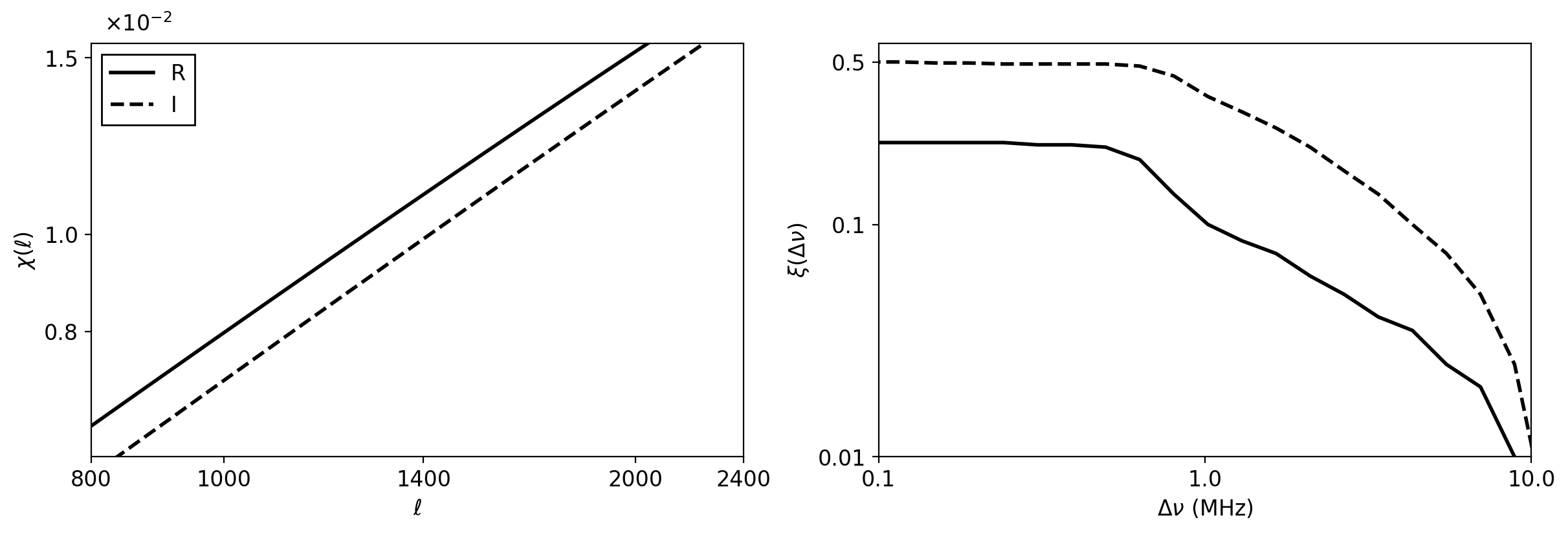}
     \caption{Left: The function $\chi(\ell)$ for the real and imaginary parts of the residual gain errors is plotted  with solid and dashed black lines, respectively, as a function of angular multipole $\ell$. Right: The function $\xi(\Delta \nu)$ for the real and imaginary parts of the residual bandpass errors is plotted as a function of frequency separation $\Delta \nu$.}
         \label{fig:chixi}
\end{figure}
In this section we use the estimates of the statistics of the residual gain- and bandpass-error and the estimation of the foreground MAPS in the target ELAIS-N1 field to analytically estimate the bias and variance in the MAPS. Figure~\ref{fig:chixi} shows the estimates of the function $\chi(\ell)$, defined in Eq.~(\ref{eq:chidef}) in the left panel for the real and imaginary parts of the residual gain errors. The right panel of the same figure shows the function $\xi(\Delta \nu)$ as estimated from the representative bandpass scan of the observation for the two complex components. Note that as stated earlier, the two polarisation components of the residual gain errors show quite similar statistics, and we use their average for this analysis.  The standard deviations of the real and imaginary parts of the residual gain errors were $4.55\%$ and $2.80\%$, respectively, and those of the residual bandpass errors were $26.20\%$ and $8.03\%$, respectively.  The frequency range of observations used here corresponds to a representative frequency of $444.0$ MHz and a redshift of $z\sim 2.2$. We estimate that at this frequency, for an integration time of $2.68$ sec and channel width of $\sim 24$ kHz, for the uGMRT $\sigma_N = 592$ mK$^2$. 

\begin{figure}[ht!]
    \centering
  \includegraphics[width=0.7\textwidth]{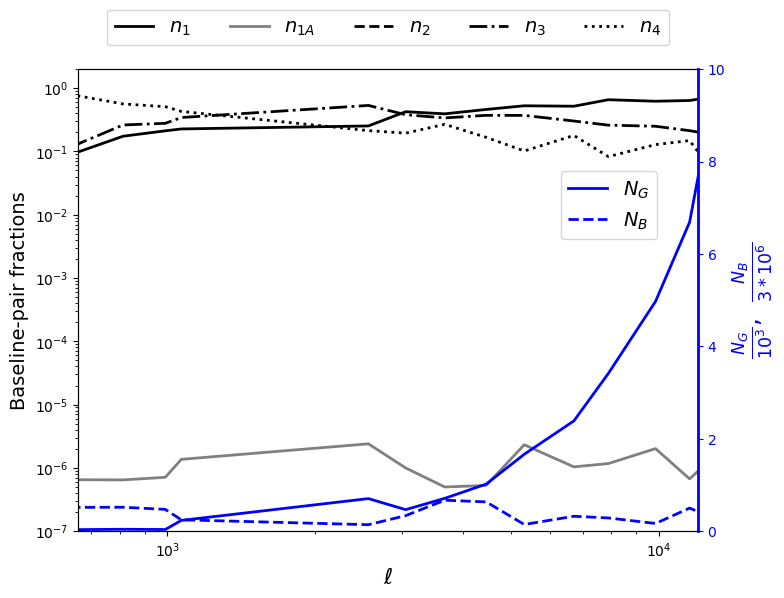}
     \caption{\bf The baseline pair fractions  $n_{i}$, the quantities $N_G$ and $N_B$ are plotted against the angular multipole $\ell$ for the observation of May 07. The left-hand axis corresponds to the baseline pair fractions  $n_{i}$ and the right-hand axis (blue) corresponds to scaled values of $N_G$ and $N_B$. Integration time and uv-grid size for the observation are kept at 2.68 sec and $0.024$ k$\lambda$ respectively.}
         \label{fig:bpfrac}
\end{figure}
The observing scans for May 06, 07 and 27 span nearly the same local sidereal time (LST) range, so we use the baselines measured  during the target observations on May 07 to estimate the baseline pair fractions  $n_{i}$, the quantities $N_G$ and $N_B$. The procedure for calculating the baseline pair fractions is discussed in detail in Paper~IV. We use a grid size of $0.024$ k$\lambda$ in the uv-plane for the CLTGE calculation. Figure~\ref{fig:bpfrac} shows the baseline pair fractions (left side vertical margin) and scaled values of $N_G$ and $N_B$ as a function of angular multipoles $\ell$ (right side vertical margin). We observe that the baseline pairs of Types~1A and 2 do not contribute much across the $\ell$ values for MAPS estimates. The Type~4 baseline pairs dominate at the shortest $\ell$ values shown here.

\begin{figure*}[ht!]
    \centering
    \includegraphics[width=0.95\textwidth]{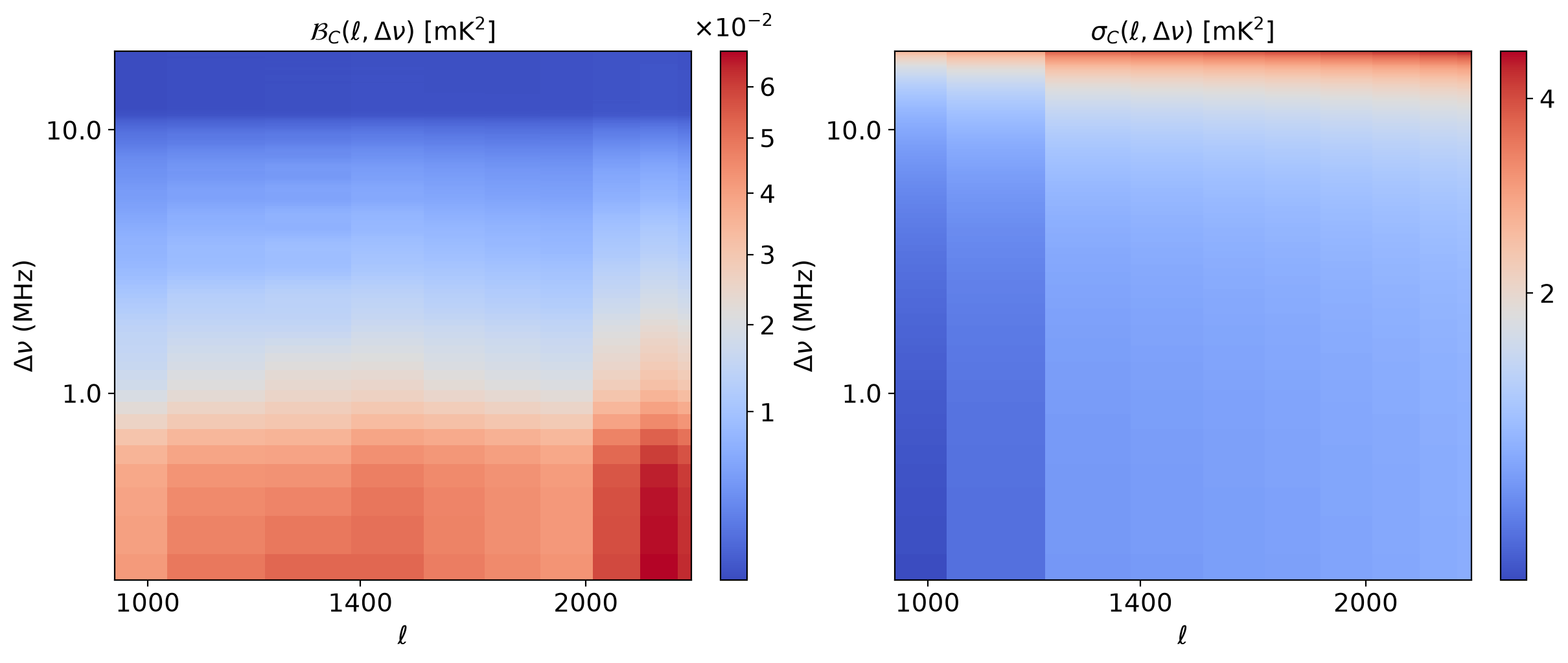}
    \caption{\bf The bias $\mathcal{B}_{C_{\ell}}$ and uncertainty $\sigma_{C_{\ell}}$  in MAPS are plotted in the $(\ell, \Delta \nu)$ plane. Both are estimated at the redshift of $z\sim 2.2$, for an integration time of $2.68$ sec and channel width of $\sim 24$ kHz. For the uGMRT, $\sigma_N = 592$ mK$^2$.} 
    \label{fig:biasvar}
\end{figure*}

\begin{figure*}[ht!]
    \centering
    \includegraphics[width=0.95\textwidth]{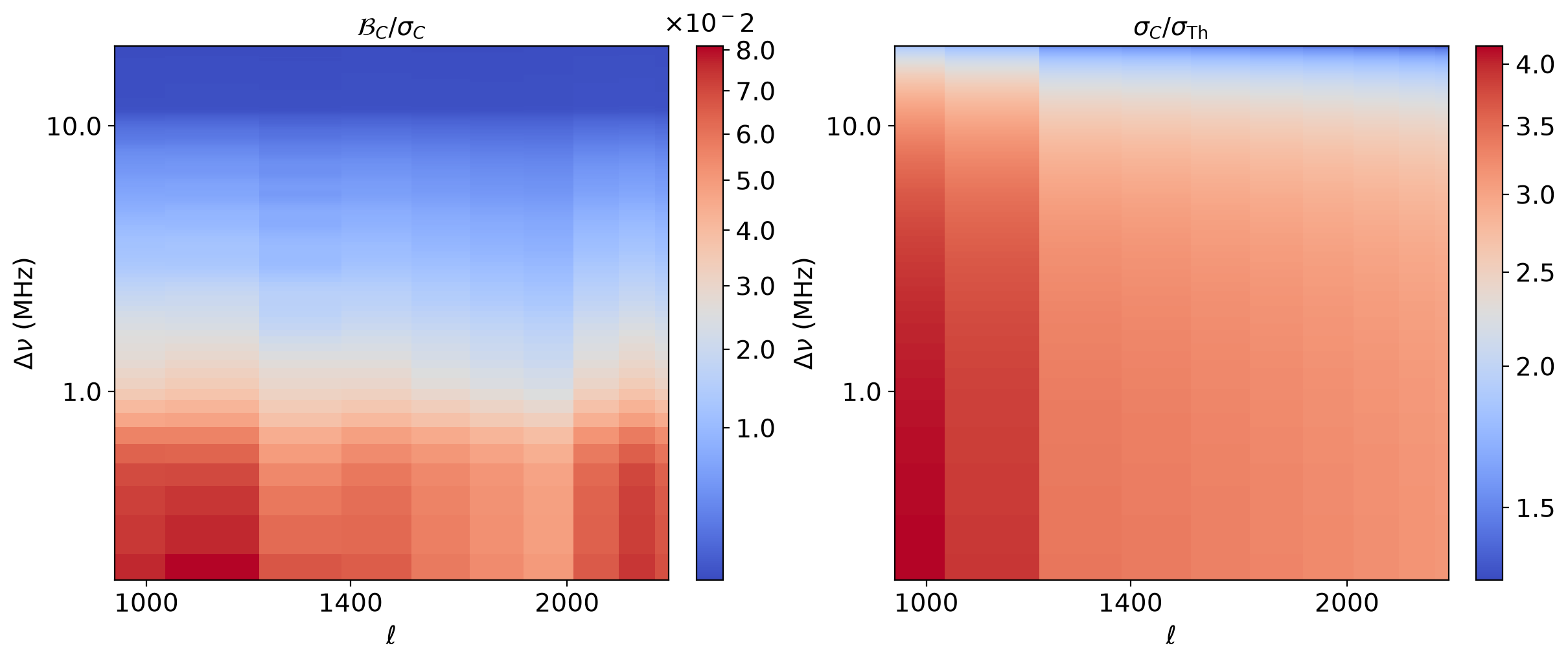}
    \caption{\bf The ratio of bias to the total uncertainty in MAPS (left) and the ratio of the total uncertainty to thermal uncertainty (right) are plotted in the $(\ell, \Delta \nu)$ plane.}
    \label{fig:bvrat}
\end{figure*}

We use the analytical framework discussed in Eqs.~(\ref{eq:bias}) and (\ref{eq:sigmaE}), respectively to calculate the bias $\mathcal{B}_{C_{\ell}}$ and variance $\sigma^2_{C_{\ell}}$  in MAPS.  The bias and the corresponding standard deviation $\sigma_{C_{\ell}}$ are shown in the left and right panels of Figure~\ref{fig:biasvar}, respectively. Figure~\ref{fig:bvrat} shows the ratio of bias to uncertainty (left) and compares the uncertainty calculated here with the thermal values of the same (right). We found that the bias in the MAPS is significantly lower than the uncertainty given by $\sigma_{C_{\ell}}$ for the particular uGMRT observation used here. The uncertainties calculated using the effect of the residual gain errors are approximately three times the thermal values. The uncertainties in the estimates are also approximately a factor of $10$ smaller than the foreground estimates; this justifies our assumption that the foreground estimates are robust.


\subsection{Bias and uncertainty in $P(k_{\perp}, k_{||})$}
\begin{figure*}[ht!]
    \centering
    \includegraphics[width=0.7\textwidth]{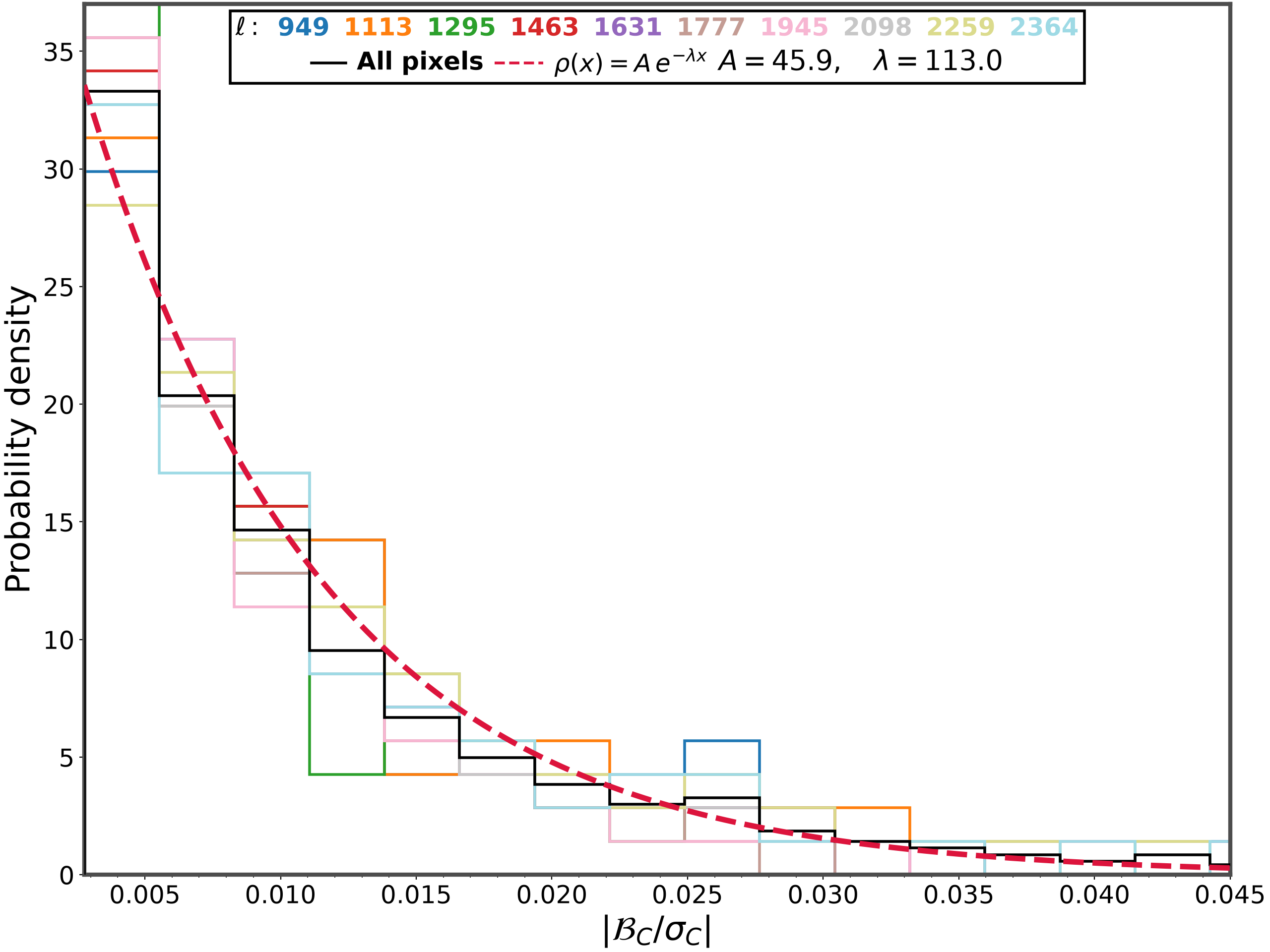}
    \caption{\bf The normalised histogram of the ratio of the bias to the standard deviation ($\mathcal{B}_{C}/\sigma_{C}$) for all pixels in $\ell-\Delta\nu$-plane is shown as a solid black line. The best-fitting exponential is shown as a red dashed curve. Coloured lines show the corresponding histograms calculated from different $\ell$ bins.}
    \label{fig:hist}
\end{figure*}

For given estimates of the bias and uncertainty in MAPS, estimating the bias and uncertainty in $P(k_{\perp}, k_{||})$ requires us to know the probability distribution of the MAPS estimates. We approach this in the following way. We assume that the estimates of MAPS have a similar underlying probability distribution for all ($\ell, \Delta \nu$) pixels. We first estimate the histogram of the ratio of the bias to the standard deviation ($\mathcal{B}_{C}/\sigma_{C}$) for all these pixels. The histogram is normalised such that it provides an estimate of the probability density. This is shown as a solid black curve  (label `All pixels') in Figure~\ref{fig:hist}. Note that here we expect that the values of the ratio are non-negative. For the subsequent Monte Carlo propagation, we adopt an exponential probability density function for the MAPS estimates as a modelling assumption. To examine whether the ratio distributions are similar across angular scales, we check the probability density of each $\ell$ bin (we have ten bins in $\ell$ here). These are shown with different colours in the same plot. We find that the probability density functions of $\mathcal{B}_{C}/\sigma_{C}$ for individual bins also follow the probability distribution from all pixels. 

\begin{figure*}[ht!]
    \centering
    \includegraphics[width=0.95\textwidth]{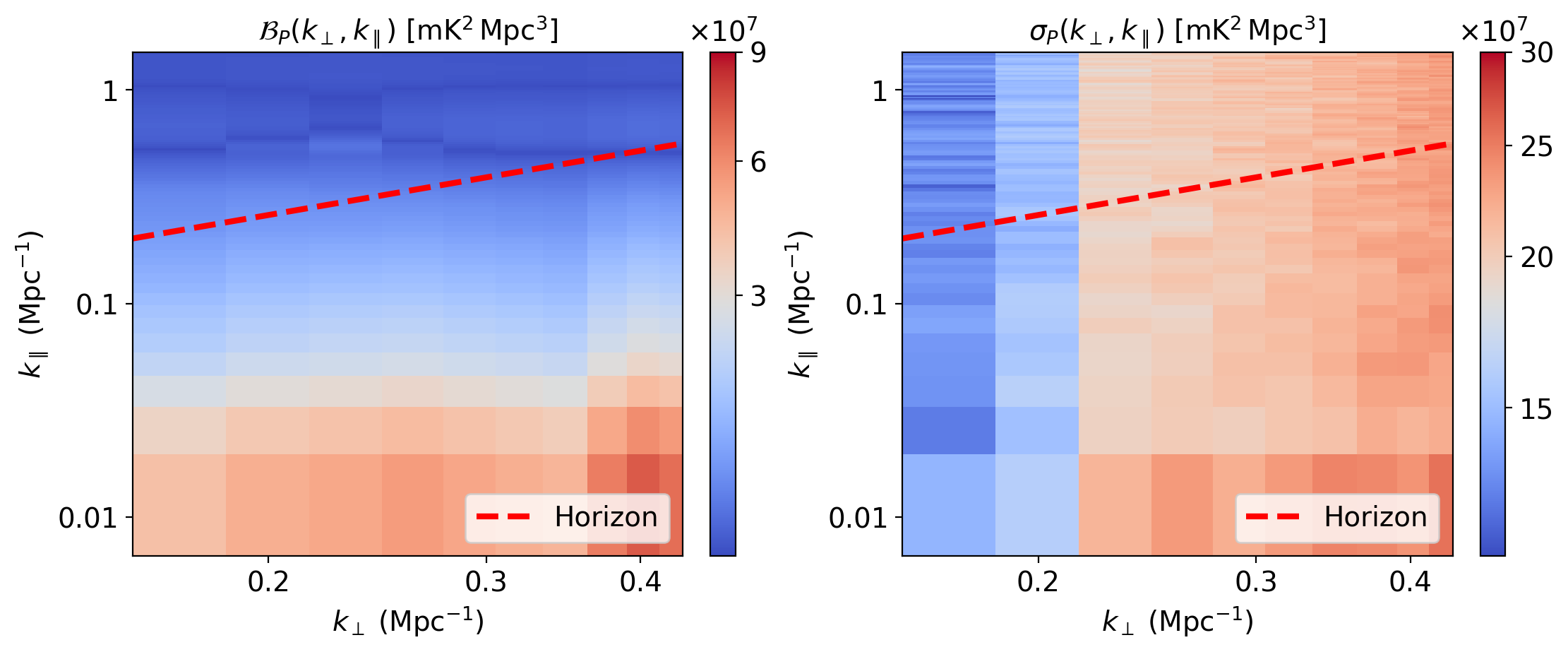}
    \caption{Estimates of the bias and standard deviation of $P(k_{\perp}, k_{||})$.}
    \label{fig:PkY}
\end{figure*}

We generate $1024$ realisations of the bias in MAPS assuming a normalised exponential distribution in each pixel in the $\ell-\Delta \nu$ plane, with the estimated bias and uncertainty values. Each of these realisations is then used to calculate the corresponding cylindrical power spectrum $P(k_{\perp}, k_{||})$ using Eq.~(\ref{eq:cl2pk}). We then calculate the mean and standard deviations of these $P(k_{\perp}, k_{||})$ realisations. These are shown in Figure~\ref{fig:PkY}. The red-dashed line in both panels corresponds to the horizon wedge limit. We observe that the bias, hence estimated in  $P(k_{\perp}, k_{||})$ is small, as expected, compared to its standard deviation. We also note that the residual gain errors contaminate the region above the wedge adding to the systematics in the observation.

\begin{figure*}[ht!]
    \centering
    \includegraphics[width=0.95\textwidth]{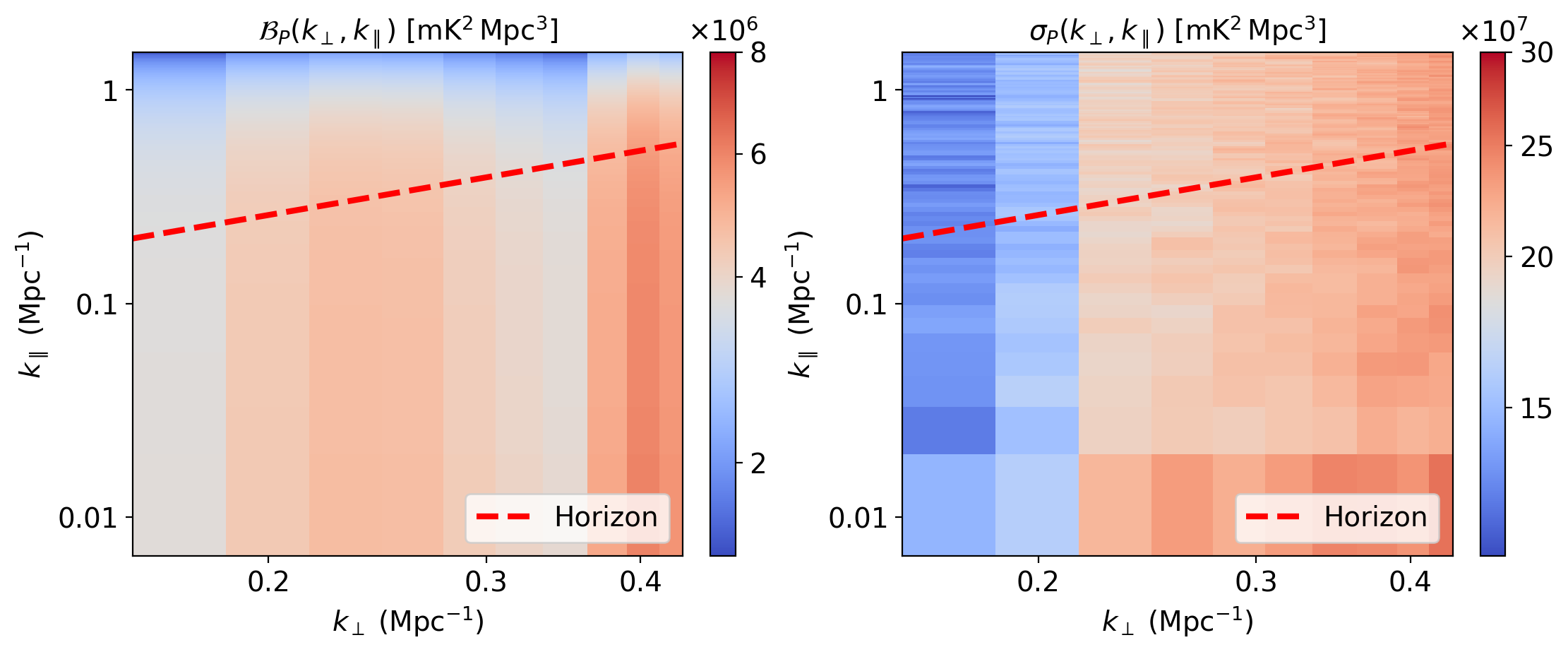}
    \caption{Same as  Figure~\ref{fig:PkY}, using Cross-TGE.}
    \label{fig:PkN}
\end{figure*}
Elahi et al. (2023) \cite{2023MNRAS.520.2094E} have addressed the correlation in the residual gain errors and have introduced a cross-correlation-based version of CLTGE (Cross-TGE), where they correlate the visibilities from two different polarisations.  Such an estimator does not have the effect of self correlation of gains in time and bandpass for antennas. Hence, the bias and variance for such estimators can be calculated from our formalism by setting $\eta_C(\tau)$ and $\xi(\Delta \nu)$ as delta functions at $\tau=0$ and $\Delta \nu=0$ respectively in Eq.~(\ref{eq:gbmod}). In Figure~\ref{fig:PkN} we show the estimates of bias and uncertainty in $P(k_{\perp}, k_{||})$  for such an estimator. We notice that Cross-TGE significantly reduces the  bias in the cylindrically averaged power spectrum; however, the excess uncertainty in the estimates of the power spectrum does not change significantly.  The method presented here, hence objectively demonstrates the advantage of using the Cross-TGE estimator and hence can be used to assess the quality of power spectrum estimators in general.

\section{Conclusion}
\label{sec:discussion}

In this work, we have combined an empirical characterisation of the residual time- and frequency-dependent calibration errors with the analytical framework developed in Paper~IV to quantify their impact on the redshifted 21-cm power spectrum for uGMRT Band~3 observations of the ELAIS-N1 field.  We note the following findings here:

\begin{itemize}
\item The analysis shows that the residual gain- and bandpass errors are, to a good approximation, Gaussian random variables with small antenna-to-antenna variation (Figures~\ref{fig:gainsummary} and \ref{fig:bpasssummary}), and this justifies the use of the analytical expressions for the bias and variance given by Eqs.~(\ref{eq:bias}) and (\ref{eq:sigmaE}). At the same time, the gain statistics reveal several non-ideal features that are relevant for future analysis. In both the time and frequency domains, we identify regions  that differ statistically  from the bulk of the data, which motivates the additional gain-statistics-based flagging introduced in Section~\ref{sec:adgainstat}. This demonstrates that the calibration tables themselves carry useful information for data quality assessment, beyond the conventional  visibility-based flagging.

\item Assessment of correlations  in residual time-dependent gain and frequency-dependent bandpass errors   provides further useful physical insight. We find that the auto-correlations in the residual calibration errors are much more pronounced in the imaginary-imaginary components than in the real-real components (Figure~\ref{fig:tcorrfun}). For the time-dependent residual gain errors, the imaginary-imaginary components remain correlated typically up to $\sim 15$ sec, whereas other components decorrelate within approximately two integration times. A similar trend is seen for the residual bandpass errors (Figure~\ref{fig:bpcorrfun}), where the imaginary component shows frequency correlation typically up to $\sim 1.5$ MHz, while the real component is confined to much smaller separations. We also find weak but non-zero cross-correlations for certain antenna/polarisation combinations for imaginary-imaginary pairs.  Note that these effects are not explicitly considered  in the simplifying assumptions of the present analytical model.  We observe that for the residual gain errors about $8\%$ (Figure~\ref{fig:tcrosscorr}) of the cross-pairs are correlated, whereas for the residual bandpass errors they remain below $\sim 2\%$ (Figure~\ref{fig:bcrosscorr}).  

\item The calibration errors are certainly specific to a given radio interferometer. Moreover, the baseline distribution of the telescope plays a significant role in the statistics of bias and uncertainty of the MAPS estimates. Our analysis shows that for the uGMRT Band~3, the bias in MAPS estimates remains smaller than the standard deviation over the full range of $\ell$ and $\Delta \nu$ considered here (Figure~\ref{fig:bvrat}). The systematic contribution to the variance is more important at lower $\ell$ and smaller $\Delta \nu$. For lower values of $\ell$ the foreground contribution is strongest, and for lower $\Delta \nu$  the residual bandpass  errors remain relatively more correlated. Motivated by the ratio histograms, we model the MAPS estimates using an exponential distribution. We use Monte Carlo error propagation to estimate the mean and uncertainty in the  cylindrical power spectrum. We note that the bias is largely confined within the foreground wedge in the $(k_{\perp}, k_{\parallel})$ plane and is significantly smaller than the uncertainties in the estimates. The uncertainty in the power spectrum, however, extends beyond the wedge to the `EoR window' (Figure~\ref{fig:PkY}). This shows that, for the uGMRT,  even when the mean level of calibration-induced contamination is modest, the corresponding uncertainty can still leak power above the wedge. As an observation is statistically only one realisation of the sky, the excess uncertainty induced by systematic effects limits the sensitivity of the telescope in the nominal ``EoR window''. The fact that the variance dominates over the bias for this observation also suggests that, for the present dataset, reducing the scatter introduced by residual calibration errors is more important than correcting a large deterministic offset.

\item We have also estimated the bias and variance using the Cross-TGE methodology discussed in the uGMRT cross-polarisation power spectrum work \cite{2023MNRAS.520.2094E}, which significantly reduces the effect of time/frequency correlations in the residual calibration error. We find that with Cross-TGE, the bias in the cylindrical power spectrum decreases further over the full range of $k_{\perp}$ and $k_{\parallel}$; however, the uncertainty remains nearly unchanged. Since the observations with the uGMRT are dominated by the uncertainty, the use of this method does not significantly change  the conclusions presented here. 
\end{itemize}
We note a few important limitations of the present work; these will be addressed in future work.

\begin{enumerate}
    \item  The present implementation uses one representative target scan and one nearby bandpass scan to characterise the residual calibration statistics. In reality, these statistics vary from scan to scan, and therefore the corresponding bias and variance should also be estimated scan-wise. Such an approach will provide a more accurate estimate of the calibration-induced systematics for a given observation. This is particularly relevant in situations where the bias may dominate over the variance, as seen in Paper~IV for the SKA1-Low case. In that regime, a scan-wise characterisation of the residual gain and bandpass errors may provide a practical route to identify and mitigate systematic effects in MAPS.    
    \item Our analysis assumes that the foreground MAPS is known accurately. Any uncertainty in the foreground estimate will introduce an additional contribution to the total uncertainty in the MAPS and the cylindrical power spectrum.   At present, the systematics are large enough that the accuracy in the foreground estimates is not the limiting factor.  This, however, needs to be addressed for more sensitive observations in the future.
    \item  Although reducing the integration time and channel width would help suppress the effect of residual time and frequency correlations, this would also increase the noise per visibility. Therefore, any future optimisation must balance the competing requirements of reduced calibration correlation and acceptable thermal noise.
    \item We observed some antenna-antenna cross correlations, particularly for the imaginary-imaginary pairs of the residual gains. These are also more prominently present for nearby antennas suggesting either cross-talk between antennas or the presence of unsubtracted low-level RFI. We plan to investigate their causes and methods for mitigating these effects in a future study.
 \end{enumerate}

Overall, the present results show that the gain solutions from an observation can be used to diagnose non-ideal calibration behaviour, guide additional flagging, and quantify the level at which residual calibration errors affect 21-cm power-spectrum measurements. The observed dominance of the imaginary-part correlations, the existence of weak cross-antenna correlations, and the leakage of variance beyond the wedge collectively highlight the need for improved calibration strategies, especially for future high-sensitivity experiments. Extending this work to a scan-wise analysis of all observing days, incorporating foreground-estimation uncertainty, and developing a formalism that includes the measured cross-correlations are the natural next steps toward a more complete treatment of calibration systematics in high-dynamic-range 21-cm observations.

\section*{ACKNOWLEDGEMENTS}
PD acknowledges discussions with Wasim Raja about various aspects of this work. PD would also like to acknowledge the research grant MATRICS (MTR/2023/000982) from SERB, India for funding a part of this work. SC would like to thank Philip Bull for useful discussions. The support and the resources provided by ‘PARAM Shivay Facility’ under the National Supercomputing Mission, Government of India at the Indian Institute of Technology, Varanasi, are gratefully acknowledged.

\section*{DATA AVAILABILITY}
This work analyses archival uGMRT observations (project codes 32\_120 and 46\_032) and generates derived analysis products. No new observations were obtained for this study.

\bibliographystyle{JHEP}
\bibliography{references.bib}

\providecommand{\href}[2]{#2}\begingroup\raggedright\begin{thebibliography}{10}

\bibitem{1985ApJ...292..371D}
M.~{Davis}, G.~{Efstathiou}, C.S.~{Frenk} and S.D.M.~{White}, \emph{{The
  evolution of large-scale structure in a universe dominated by cold dark
  matter}}, \href{https://doi.org/10.1086/163168}{\emph{\apj} {\bfseries 292}
  (1985) 371}.

\bibitem{1990MNRAS.247..510S}
D.~{Scott} and M.J.~{Rees}, \emph{{The 21-cm line at high redshift: a
  diagnostic for the origin of large scale structure}}, {\emph{\mnras}
  {\bfseries 247} (1990) 510}.

\bibitem{1993MNRAS.265..101S}
K.~{Subramanian} and T.~{Padmanabhan}, \emph{{Neutral Hydrogen at High
  Redshifts as a Probe of Structure Formation - Part One - Post-Cobe Analysis
  of CDM and HDM Models}},
  \href{https://doi.org/10.1093/mnras/265.1.101}{\emph{\mnras} {\bfseries 265}
  (1993) 101}.

\bibitem{1995MNRAS.272..544K}
A.~{Kumar}, T.~{Padmanabhan} and K.~{Subramanian}, \emph{{Neutral hydrogen at
  high redshifts as a probe of structure formation - II. Line profile of a
  protocluster}}, \href{https://doi.org/10.1093/mnras/272.3.544}{\emph{\mnras}
  {\bfseries 272} (1995) 544}.

\bibitem{1997ApJ...475..429M}
P.~{Madau}, A.~{Meiksin} and M.J.~{Rees}, \emph{{21 Centimeter Tomography of
  the Intergalactic Medium at High Redshift}},
  \href{https://doi.org/10.1086/303549}{\emph{\apj} {\bfseries 475} (1997) 429}
  [\href{https://arxiv.org/abs/astro-ph/9608010}{{\ttfamily
  astro-ph/9608010}}].

\bibitem{2001JApA...22...21B}
S.~{Bharadwaj}, B.B.~{Nath} and S.K.~{Sethi}, \emph{{Using HI to probe large
  scale structures at z{\ensuremath{\sim}}3}},
  \href{https://doi.org/10.1007/BF02933588}{\emph{Journal of Astrophysics and
  Astronomy} {\bfseries 22} (2001) 21}
  [\href{https://arxiv.org/abs/astro-ph/0003200}{{\ttfamily
  astro-ph/0003200}}].

\bibitem{2010ARA&A..48..127M}
M.F.~{Morales} and J.S.B.~{Wyithe}, \emph{{Reionization and Cosmology with
  21-cm Fluctuations}},
  \href{https://doi.org/10.1146/annurev-astro-081309-130936}{\emph{\araa}
  {\bfseries 48} (2010) 127} [\href{https://arxiv.org/abs/0910.3010}{{\ttfamily
  0910.3010}}].

\bibitem{2004ApJ...608..622Z}
M.~{Zaldarriaga}, S.R.~{Furlanetto} and L.~{Hernquist}, \emph{{21 Centimeter
  Fluctuations from Cosmic Gas at High Redshifts}},
  \href{https://doi.org/10.1086/386327}{\emph{\apj} {\bfseries 608} (2004) 622}
  [\href{https://arxiv.org/abs/astro-ph/0311514}{{\ttfamily
  astro-ph/0311514}}].

\bibitem{2004ApJ...613....1F}
S.R.~{Furlanetto}, M.~{Zaldarriaga} and L.~{Hernquist}, \emph{{The Growth of H
  II Regions During Reionization}},
  \href{https://doi.org/10.1086/423025}{\emph{\apj} {\bfseries 613} (2004) 1}
  [\href{https://arxiv.org/abs/astro-ph/0403697}{{\ttfamily
  astro-ph/0403697}}].

\bibitem{2005ApJ...625..575S}
M.G.~{Santos}, A.~{Cooray} and L.~{Knox}, \emph{{Multifrequency Analysis of 21
  Centimeter Fluctuations from the Era of Reionization}},
  \href{https://doi.org/10.1086/429857}{\emph{\apj} {\bfseries 625} (2005) 575}
  [\href{https://arxiv.org/abs/astro-ph/0408515}{{\ttfamily
  astro-ph/0408515}}].

\bibitem{2006PhR...433..181F}
S.R.~{Furlanetto}, S.P.~{Oh} and F.H.~{Briggs}, \emph{{Cosmology at low
  frequencies: The 21 cm transition and the high-redshift Universe}},
  \href{https://doi.org/10.1016/j.physrep.2006.08.002}{\emph{\physrep}
  {\bfseries 433} (2006) 181}
  [\href{https://arxiv.org/abs/astro-ph/0608032}{{\ttfamily
  astro-ph/0608032}}].

\bibitem{2001JApA...22..293B}
S.~{Bharadwaj} and S.K.~{Sethi}, \emph{{HI Fluctuations at Large Redshifts:
  I--Visibility correlation}},
  \href{https://doi.org/10.1007/BF02702273}{\emph{Journal of Astrophysics and
  Astronomy} {\bfseries 22} (2001) 293}
  [\href{https://arxiv.org/abs/astro-ph/0203269}{{\ttfamily
  astro-ph/0203269}}].

\bibitem{2004ApJ...615....7M}
M.F.~{Morales} and J.~{Hewitt}, \emph{{Toward Epoch of Reionization
  Measurements with Wide-Field Radio Observations}},
  \href{https://doi.org/10.1086/424437}{\emph{\apj} {\bfseries 615} (2004) 7}
  [\href{https://arxiv.org/abs/astro-ph/0312437}{{\ttfamily
  astro-ph/0312437}}].

\bibitem{2005MNRAS.356.1519B}
S.~{Bharadwaj} and S.S.~{Ali}, \emph{{On using visibility correlations to probe
  the HI distribution from the dark ages to the present epoch - I. Formalism
  and the expected signal}},
  \href{https://doi.org/10.1111/j.1365-2966.2004.08604.x}{\emph{\mnras}
  {\bfseries 356} (2005) 1519}
  [\href{https://arxiv.org/abs/astro-ph/0406676}{{\ttfamily
  astro-ph/0406676}}].

\bibitem{2007MNRAS.378..119D}
K.K.~{Datta}, T.R.~{Choudhury} and S.~{Bharadwaj}, \emph{{The multifrequency
  angular power spectrum of the epoch of reionization 21-cm signal}},
  \href{https://doi.org/10.1111/j.1365-2966.2007.11747.x}{\emph{\mnras}
  {\bfseries 378} (2007) 119}
  [\href{https://arxiv.org/abs/astro-ph/0605546}{{\ttfamily
  astro-ph/0605546}}].

\bibitem{1999A&A...345..380S}
P.A.~{Shaver}, R.A.~{Windhorst}, P.~{Madau} and A.G.~{de Bruyn}, \emph{{Can the
  reionization epoch be detected as a global signature in the cosmic
  background?}}, {\emph{\aap} {\bfseries 345} (1999) 380}
  [\href{https://arxiv.org/abs/astro-ph/9901320}{{\ttfamily
  astro-ph/9901320}}].

\bibitem{2002ApJ...564..576D}
T.~{Di Matteo}, R.~{Perna}, T.~{Abel} and M.J.~{Rees}, \emph{{Radio Foregrounds
  for the 21 Centimeter Tomography of the Neutral Intergalactic Medium at High
  Redshifts}}, \href{https://doi.org/10.1086/324293}{\emph{\apj} {\bfseries
  564} (2002) 576} [\href{https://arxiv.org/abs/astro-ph/0109241}{{\ttfamily
  astro-ph/0109241}}].

\bibitem{2004MNRAS.355.1053D}
T.~{Di Matteo}, B.~{Ciardi} and F.~{Miniati}, \emph{{The 21-cm emission from
  the reionization epoch: extended and point source foregrounds}},
  \href{https://doi.org/10.1111/j.1365-2966.2004.08443.x}{\emph{\mnras}
  {\bfseries 355} (2004) 1053}
  [\href{https://arxiv.org/abs/astro-ph/0402322}{{\ttfamily
  astro-ph/0402322}}].

\bibitem{2008MNRAS.385.2166A}
S.S.~{Ali}, S.~{Bharadwaj} and J.N.~{Chengalur}, \emph{{Foregrounds for
  redshifted 21-cm studies of reionization: Giant Meter Wave Radio Telescope
  153-MHz observations}},
  \href{https://doi.org/10.1111/j.1365-2966.2008.12984.x}{\emph{\mnras}
  {\bfseries 385} (2008) 2166}
  [\href{https://arxiv.org/abs/0801.2424}{{\ttfamily 0801.2424}}].

\bibitem{2008MNRAS.389.1319J}
V.~{Jeli{\'c}}, S.~{Zaroubi}, P.~{Labropoulos}, R.M.~{Thomas}, G.~{Bernardi},
  M.A.~{Brentjens} et~al., \emph{{Foreground simulations for the LOFAR-epoch of
  reionization experiment}},
  \href{https://doi.org/10.1111/j.1365-2966.2008.13634.x}{\emph{\mnras}
  {\bfseries 389} (2008) 1319}
  [\href{https://arxiv.org/abs/0804.1130}{{\ttfamily 0804.1130}}].

\bibitem{2012MNRAS.426.3295G}
A.~{Ghosh}, J.~{Prasad}, S.~{Bharadwaj}, S.S.~{Ali} and J.N.~{Chengalur},
  \emph{{Characterizing foreground for redshifted 21 cm radiation: 150 MHz
  Giant Metrewave Radio Telescope observations}},
  \href{https://doi.org/10.1111/j.1365-2966.2012.21889.x}{\emph{\mnras}
  {\bfseries 426} (2012) 3295}
  [\href{https://arxiv.org/abs/1208.1617}{{\ttfamily 1208.1617}}].

\bibitem{Bernardi}
G.~{Bernardi}, A.G.~{de Bruyn}, G.~{Harker}, M.A.~{Brentjens}, B.~{Ciardi},
  V.~{Jeli{\'c}} et~al., \emph{{Foregrounds for observations of the
  cosmological 21 cm line. II. Westerbork observations of the fields around 3C
  196 and the North Celestial Pole}},
  \href{https://doi.org/10.1051/0004-6361/200913420}{\emph{\aap} {\bfseries
  522} (2010) A67} [\href{https://arxiv.org/abs/1002.4177}{{\ttfamily
  1002.4177}}].

\bibitem{2006ApJ...648..767M}
M.F.~{Morales}, J.D.~{Bowman} and J.N.~{Hewitt}, \emph{{Improving Foreground
  Subtraction in Statistical Observations of 21 cm Emission from the Epoch of
  Reionization}}, \href{https://doi.org/10.1086/506135}{\emph{\apj} {\bfseries
  648} (2006) 767} [\href{https://arxiv.org/abs/astro-ph/0510027}{{\ttfamily
  astro-ph/0510027}}].

\bibitem{2009ApJ...695..183B}
J.D.~{Bowman}, M.F.~{Morales} and J.N.~{Hewitt}, \emph{{Foreground
  Contamination in Interferometric Measurements of the Redshifted 21 cm Power
  Spectrum}}, \href{https://doi.org/10.1088/0004-637X/695/1/183}{\emph{\apj}
  {\bfseries 695} (2009) 183}
  [\href{https://arxiv.org/abs/0807.3956}{{\ttfamily 0807.3956}}].

\bibitem{2017NewA...57...94C}
S.~{Choudhuri}, N.~{Roy}, S.~{Bharadwaj}, S.~{Saiyad Ali}, A.~{Ghosh} and
  P.~{Dutta}, \emph{{Validating a novel angular power spectrum estimator using
  simulated low frequency radio-interferometric data}},
  \href{https://doi.org/10.1016/j.newast.2017.06.010}{\emph{\na} {\bfseries 57}
  (2017) 94} [\href{https://arxiv.org/abs/1706.09033}{{\ttfamily 1706.09033}}].

\bibitem{2010ApJ...724..526D}
A.~{Datta}, J.D.~{Bowman} and C.L.~{Carilli}, \emph{{Bright Source Subtraction
  Requirements for Redshifted 21 cm Measurements}},
  \href{https://doi.org/10.1088/0004-637X/724/1/526}{\emph{\apj} {\bfseries
  724} (2010) 526} [\href{https://arxiv.org/abs/1005.4071}{{\ttfamily
  1005.4071}}].

\bibitem{2012ApJ...752..137M}
M.F.~{Morales}, B.~{Hazelton}, I.~{Sullivan} and A.~{Beardsley}, \emph{{Four
  Fundamental Foreground Power Spectrum Shapes for 21 cm Cosmology
  Observations}},
  \href{https://doi.org/10.1088/0004-637X/752/2/137}{\emph{\apj} {\bfseries
  752} (2012) 137} [\href{https://arxiv.org/abs/1202.3830}{{\ttfamily
  1202.3830}}].

\bibitem{2012ApJ...745..176V}
H.~{Vedantham}, N.~{Udaya Shankar} and R.~{Subrahmanyan}, \emph{{Imaging the
  Epoch of Reionization: Limitations from Foreground Confusion and Imaging
  Algorithms}}, \href{https://doi.org/10.1088/0004-637X/745/2/176}{\emph{\apj}
  {\bfseries 745} (2012) 176}
  [\href{https://arxiv.org/abs/1106.1297}{{\ttfamily 1106.1297}}].

\bibitem{2012ApJ...756..165P}
A.R.~{Parsons}, J.C.~{Pober}, J.E.~{Aguirre}, C.L.~{Carilli}, D.C.~{Jacobs} and
  D.F.~{Moore}, \emph{{A Per-baseline, Delay-spectrum Technique for Accessing
  the 21 cm Cosmic Reionization Signature}},
  \href{https://doi.org/10.1088/0004-637X/756/2/165}{\emph{\apj} {\bfseries
  756} (2012) 165} [\href{https://arxiv.org/abs/1204.4749}{{\ttfamily
  1204.4749}}].

\bibitem{2013ApJ...776....6T}
N.~{Thyagarajan}, N.~{Udaya Shankar}, R.~{Subrahmanyan}, W.~{Arcus},
  G.~{Bernardi}, J.D.~{Bowman} et~al., \emph{{A Study of Fundamental
  Limitations to Statistical Detection of Redshifted H I from the Epoch of
  Reionization}}, \href{https://doi.org/10.1088/0004-637X/776/1/6}{\emph{\apj}
  {\bfseries 776} (2013) 6} [\href{https://arxiv.org/abs/1308.0565}{{\ttfamily
  1308.0565}}].

\bibitem{2014PhRvD..90b3018L}
A.~{Liu}, A.R.~{Parsons} and C.M.~{Trott}, \emph{{Epoch of reionization window.
  I. Mathematical formalism}},
  \href{https://doi.org/10.1103/PhysRevD.90.023018}{\emph{\prd} {\bfseries 90}
  (2014) 023018} [\href{https://arxiv.org/abs/1404.2596}{{\ttfamily
  1404.2596}}].

\bibitem{2014PhRvD..90b3019L}
A.~{Liu}, A.R.~{Parsons} and C.M.~{Trott}, \emph{{Epoch of reionization window.
  II. Statistical methods for foreground wedge reduction}},
  \href{https://doi.org/10.1103/PhysRevD.90.023019}{\emph{\prd} {\bfseries 90}
  (2014) 023019} [\href{https://arxiv.org/abs/1404.4372}{{\ttfamily
  1404.4372}}].

\bibitem{2016ApJ...818..139T}
C.M.~{Trott}, B.~{Pindor}, P.~{Procopio}, R.B.~{Wayth}, D.A.~{Mitchell},
  B.~{McKinley} et~al., \emph{{CHIPS: The Cosmological H I Power Spectrum
  Estimator}}, \href{https://doi.org/10.3847/0004-637X/818/2/139}{\emph{\apj}
  {\bfseries 818} (2016) 139}
  [\href{https://arxiv.org/abs/1601.02073}{{\ttfamily 1601.02073}}].

\bibitem{2022MNRAS.516.2851P}
S.~{Pal}, K.M.A.~{Elahi}, S.~{Bharadwaj}, S.S.~{Ali}, S.~{Choudhuri},
  A.~{Ghosh} et~al., \emph{{Towards 21-cm intensity mapping at z = 2.28 with
  uGMRT using the tapered gridded estimator I: Foreground avoidance}},
  \href{https://doi.org/10.1093/mnras/stac2419}{\emph{\mnras} {\bfseries 516}
  (2022) 2851} [\href{https://arxiv.org/abs/2208.11063}{{\ttfamily
  2208.11063}}].

\bibitem{2016MNRAS.463.4093C}
S.~{Choudhuri}, S.~{Bharadwaj}, S.~{Chatterjee}, S.S.~{Ali}, N.~{Roy} and
  A.~{Ghosh}, \emph{{The visibility-based tapered gridded estimator (TGE) for
  the redshifted 21-cm power spectrum}},
  \href{https://doi.org/10.1093/mnras/stw2254}{\emph{\mnras} {\bfseries 463}
  (2016) 4093} [\href{https://arxiv.org/abs/1609.01732}{{\ttfamily
  1609.01732}}].

\bibitem{2019MNRAS.483.3910C}
S.~{Choudhuri}, P.~{Dutta} and S.~{Bharadwaj}, \emph{{An Image-based Tapered
  Gridded Estimator (ITGE) for the angular power spectrum}},
  \href{https://doi.org/10.1093/mnras/sty3429}{\emph{\mnras} {\bfseries 483}
  (2019) 3910} [\href{https://arxiv.org/abs/1812.06995}{{\ttfamily
  1812.06995}}].

\bibitem{2019MNRAS.483.5694B}
S.~{Bharadwaj}, S.~{Pal}, S.~{Choudhuri} and P.~{Dutta}, \emph{{A Tapered
  Gridded Estimator (TGE) for the multifrequency angular power spectrum (MAPS)
  and the cosmological H I 21-cm power spectrum}},
  \href{https://doi.org/10.1093/mnras/sty3501}{\emph{\mnras} {\bfseries 483}
  (2019) 5694} [\href{https://arxiv.org/abs/1812.08801}{{\ttfamily
  1812.08801}}].

\bibitem{2021MNRAS.507.5310A}
U.~{Arora} and P.~{Dutta}, \emph{{Combined lensed estimator to probe the
  post-reionization H I power spectrum}},
  \href{https://doi.org/10.1093/mnras/stab2480}{\emph{\mnras} {\bfseries 507}
  (2021) 5310} [\href{https://arxiv.org/abs/2108.12677}{{\ttfamily
  2108.12677}}].

\bibitem{2023MNRAS.520.2094E}
K.M.A.~{Elahi}, S.~{Bharadwaj}, A.~{Ghosh}, S.~{Pal}, S.S.~{Ali},
  S.~{Choudhuri} et~al., \emph{{Towards 21-cm intensity mapping at z = 2.28
  with uGMRT using the tapered gridded estimator - II. Cross-polarization power
  spectrum}}, \href{https://doi.org/10.1093/mnras/stad191}{\emph{mnras}
  {\bfseries 520} (2023) 2094}
  [\href{https://arxiv.org/abs/2301.06677}{{\ttfamily 2301.06677}}].

\bibitem{2013MNRAS.433..639P}
G.~{Paciga}, J.G.~{Albert}, K.~{Bandura}, T.-C.~{Chang}, Y.~{Gupta},
  C.~{Hirata} et~al., \emph{{A simulation-calibrated limit on the H I power
  spectrum from the GMRT Epoch of Reionization experiment}},
  \href{https://doi.org/10.1093/mnras/stt753}{\emph{\mnras} {\bfseries 433}
  (2013) 639} [\href{https://arxiv.org/abs/1301.5906}{{\ttfamily 1301.5906}}].

\bibitem{2019ApJ...884....1B}
N.~{Barry}, M.~{Wilensky}, C.M.~{Trott}, B.~{Pindor}, A.P.~{Beardsley},
  B.J.~{Hazelton} et~al., \emph{{Improving the Epoch of Reionization Power
  Spectrum Results from Murchison Widefield Array Season 1 Observations}},
  \href{https://doi.org/10.3847/1538-4357/ab40a8}{\emph{\apj} {\bfseries 884}
  (2019) 1} [\href{https://arxiv.org/abs/1909.00561}{{\ttfamily 1909.00561}}].

\bibitem{2020MNRAS.493.1662M}
F.G.~{Mertens}, M.~{Mevius}, L.V.E.~{Koopmans}, A.R.~{Offringa}, G.~{Mellema},
  S.~{Zaroubi} et~al., \emph{{Improved upper limits on the 21 cm signal power
  spectrum of neutral hydrogen at z {\ensuremath{\approx}} 9.1 from LOFAR}},
  \href{https://doi.org/10.1093/mnras/staa327}{\emph{\mnras} {\bfseries 493}
  (2020) 1662} [\href{https://arxiv.org/abs/2002.07196}{{\ttfamily
  2002.07196}}].

\bibitem{2025A&A...698A.186M}
F.G.~{Mertens}, M.~{Mevius}, L.V.E.~{Koopmans}, A.R.~{Offringa}, S.~{Zaroubi},
  A.~{Acharya} et~al., \emph{{Deeper multi-redshift upper limits on the epoch
  of reionisation 21 cm signal power spectrum from LOFAR between z = 8.3 and z
  = 10.1}}, \href{https://doi.org/10.1051/0004-6361/202554158}{\emph{aap}
  {\bfseries 698} (2025) A186}
  [\href{https://arxiv.org/abs/2503.05576}{{\ttfamily 2503.05576}}].

\bibitem{2025MNRAS.544.3617S}
R.~{Sagar}, A.~{Datta}, A.~{Chakraborty}, N.~{Roy}, A.~{Sinha}, A.~{Mazumder}
  et~al., \emph{{ELAIS-N1 deep field uGMRT Band-2: constraints on diffuse
  Galactic synchrotron emission power spectrum}},
  \href{https://doi.org/10.1093/mnras/staf1922}{\emph{\mnras} {\bfseries 544}
  (2025) 3617} [\href{https://arxiv.org/abs/2511.02375}{{\ttfamily
  2511.02375}}].

\bibitem{1984ARA&A..22...97P}
T.J.~{Pearson} and A.C.S.~{Readhead}, \emph{{Image Formation by
  Self-Calibration in Radio Astronomy}},
  \href{https://doi.org/10.1146/annurev.aa.22.090184.000525}{\emph{\araa}
  {\bfseries 22} (1984) 97}.

\bibitem{1992ExA.....2..203W}
M.H.~{Wieringa}, \emph{{An investigation of the telescope based calibration
  methods `redundancy' and `self-cal'}},
  \href{https://doi.org/10.1007/BF00420576}{\emph{Experimental Astronomy}
  {\bfseries 2} (1992) 203}.

\bibitem{2007ITSP...55.4497V}
S.~{van der Tol}, B.D.~{Jeffs} and A.J.~{van der Veen}, \emph{{Self-Calibration
  for the LOFAR Radio Astronomical Array}},
  \href{https://doi.org/10.1109/TSP.2007.896243}{\emph{IEEE Transactions on
  Signal Processing} {\bfseries 55} (2007) 4497}.

\bibitem{2009ITSP...57.3512W}
S.J.~{Wijnholds} and A.-J.~{van der Veen}, \emph{{Multisource Self-Calibration
  for Sensor Arrays}},
  \href{https://doi.org/10.1109/TSP.2009.2022894}{\emph{IEEE Transactions on
  Signal Processing} {\bfseries 57} (2009) 3512}
  [\href{https://arxiv.org/abs/1003.1892}{{\ttfamily 1003.1892}}].

\bibitem{2016MNRAS.461.3135B}
N.~{Barry}, B.~{Hazelton}, I.~{Sullivan}, M.F.~{Morales} and J.C.~{Pober},
  \emph{{Calibration requirements for detecting the 21 cm epoch of reionization
  power spectrum and implications for the SKA}},
  \href{https://doi.org/10.1093/mnras/stw1380}{\emph{\mnras} {\bfseries 461}
  (2016) 3135} [\href{https://arxiv.org/abs/1603.00607}{{\ttfamily
  1603.00607}}].

\bibitem{2016MNRAS.463.4317P}
A.H.~{Patil}, S.~{Yatawatta}, S.~{Zaroubi}, L.V.E.~{Koopmans}, A.G.~{de Bruyn},
  V.~{Jeli{\'c}} et~al., \emph{{Systematic biases in low-frequency radio
  interferometric data due to calibration: the LOFAR-EoR case}},
  \href{https://doi.org/10.1093/mnras/stw2277}{\emph{\mnras} {\bfseries 463}
  (2016) 4317} [\href{https://arxiv.org/abs/1605.07619}{{\ttfamily
  1605.07619}}].

\bibitem{2025JCAP...10..035T}
A.~{Tripathi}, A.~{Datta}, A.~{Mazumder} and S.~{Majumdar}, \emph{{Impact of
  calibration and position errors on astrophysical parameters of the HI 21cm
  signal}}, \href{https://doi.org/10.1088/1475-7516/2025/10/035}{\emph{\jcap}
  {\bfseries 2025} (2025) 035}
  [\href{https://arxiv.org/abs/2502.20962}{{\ttfamily 2502.20962}}].

\bibitem{2018MNRAS.478.1484G}
B.K.~{Gehlot}, L.V.E.~{Koopmans}, A.G.~{de Bruyn}, S.~{Zaroubi},
  M.A.~{Brentjens}, K.M.B.~{Asad} et~al., \emph{{Wide-field LOFAR-LBA
  power-spectra analyses: impact of calibration, polarization leakage, and
  ionosphere}}, \href{https://doi.org/10.1093/mnras/sty1095}{\emph{\mnras}
  {\bfseries 478} (2018) 1484}
  [\href{https://arxiv.org/abs/1709.07727}{{\ttfamily 1709.07727}}].

\bibitem{2015MNRAS.451.3709A}
K.M.B.~{Asad}, L.V.E.~{Koopmans}, V.~{Jeli{\'c}}, V.N.~{Pandey}, A.~{Ghosh},
  F.B.~{Abdalla} et~al., \emph{{Polarization leakage in epoch of reionization
  windows - I. Low Frequency Array observations of the 3C196 field}},
  \href{https://doi.org/10.1093/mnras/stv1107}{\emph{\mnras} {\bfseries 451}
  (2015) 3709} [\href{https://arxiv.org/abs/1503.01644}{{\ttfamily
  1503.01644}}].

\bibitem{2016JApA...37...35A}
S.S.~{Ali}, S.~{Bharadwaj}, S.~{Choudhuri}, A.~{Ghosh} and N.~{Roy},
  \emph{{Prospects of Measuring the Angular Power Spectrum of the Diffuse
  Galactic Synchrotron Emission with SKA1 Low}},
  \href{https://doi.org/10.1007/s12036-016-9413-x}{\emph{Journal of
  Astrophysics and Astronomy} {\bfseries 37} (2016) 35}
  [\href{https://arxiv.org/abs/1610.08184}{{\ttfamily 1610.08184}}].

\bibitem{2018MNRAS.476.3051A}
K.M.B.~{Asad}, L.V.E.~{Koopmans}, V.~{Jeli{\'c}}, A.G.~{de Bruyn},
  V.N.~{Pandey} and B.K.~{Gehlot}, \emph{{Polarization leakage in epoch of
  reionization windows - III. Wide-field effects of narrow-field arrays}},
  \href{https://doi.org/10.1093/mnras/sty258}{\emph{\mnras} {\bfseries 476}
  (2018) 3051} [\href{https://arxiv.org/abs/1706.00875}{{\ttfamily
  1706.00875}}].

\bibitem{2015MNRAS.453..925V}
H.K.~{Vedantham} and L.V.E.~{Koopmans}, \emph{{Scintillation noise in widefield
  radio interferometry}},
  \href{https://doi.org/10.1093/mnras/stv1594}{\emph{\mnras} {\bfseries 453}
  (2015) 925} [\href{https://arxiv.org/abs/1412.1420}{{\ttfamily 1412.1420}}].

\bibitem{2016MNRAS.458.3099V}
H.K.~{Vedantham} and L.V.E.~{Koopmans}, \emph{{Scintillation noise power
  spectrum and its impact on high-redshift 21-cm observations}},
  \href{https://doi.org/10.1093/mnras/stw443}{\emph{\mnras} {\bfseries 458}
  (2016) 3099} [\href{https://arxiv.org/abs/1512.00159}{{\ttfamily
  1512.00159}}].

\bibitem{2016RaSc...51..927M}
M.~{Mevius}, S.~{van der Tol}, V.N.~{Pandey}, H.K.~{Vedantham},
  M.A.~{Brentjens}, A.G.~{de Bruyn} et~al., \emph{{Probing ionospheric
  structures using the LOFAR radio telescope}},
  \href{https://doi.org/10.1002/2016RS006028}{\emph{Radio Science} {\bfseries
  51} (2016) 927} [\href{https://arxiv.org/abs/1606.04683}{{\ttfamily
  1606.04683}}].

\bibitem{2017MNRAS.470.1849E}
A.~{Ewall-Wice}, J.S.~{Dillon}, A.~{Liu} and J.~{Hewitt}, \emph{{The impact of
  modelling errors on interferometer calibration for 21 cm power spectra}},
  \href{https://doi.org/10.1093/mnras/stx1221}{\emph{\mnras} {\bfseries 470}
  (2017) 1849} [\href{https://arxiv.org/abs/1610.02689}{{\ttfamily
  1610.02689}}].

\bibitem{2019ApJ...875...70B}
R.~{Byrne}, M.F.~{Morales}, B.~{Hazelton}, W.~{Li}, N.~{Barry},
  A.P.~{Beardsley} et~al., \emph{{Fundamental Limitations on the Calibration of
  Redundant 21 cm Cosmology Instruments and Implications for HERA and the
  SKA}}, \href{https://doi.org/10.3847/1538-4357/ab107d}{\emph{\apj} {\bfseries
  875} (2019) 70} [\href{https://arxiv.org/abs/1811.01378}{{\ttfamily
  1811.01378}}].

\bibitem{2018AJ....156..285J}
R.C.~{Joseph}, C.M.~{Trott} and R.B.~{Wayth}, \emph{{The Bias and Uncertainty
  of Redundant and Sky-based Calibration Under Realistic Sky and Telescope
  Conditions}}, \href{https://doi.org/10.3847/1538-3881/aaec0b}{\emph{\aj}
  {\bfseries 156} (2018) 285}
  [\href{https://arxiv.org/abs/1810.11237}{{\ttfamily 1810.11237}}].

\bibitem{2019MNRAS.487..537O}
N.~{Orosz}, J.S.~{Dillon}, A.~{Ewall-Wice}, A.R.~{Parsons} and
  N.~{Thyagarajan}, \emph{{Mitigating the effects of antenna-to-antenna
  variation on redundant-baseline calibration for 21 cm cosmology}},
  \href{https://doi.org/10.1093/mnras/stz1287}{\emph{\mnras} {\bfseries 487}
  (2019) 537} [\href{https://arxiv.org/abs/1809.09728}{{\ttfamily
  1809.09728}}].

\bibitem{2022ApJ...941..207K}
H.~{Kim}, B.D.~{Nhan}, J.N.~{Hewitt}, N.S.~{Kern}, J.S.~{Dillon}, E.~{de Lera
  Acedo} et~al., \emph{{The Impact of Beam Variations on Power Spectrum
  Estimation for 21 cm Cosmology. I. Simulations of Foreground Contamination
  for HERA}}, \href{https://doi.org/10.3847/1538-4357/ac9eaf}{\emph{apj}
  {\bfseries 941} (2022) 207}
  [\href{https://arxiv.org/abs/2210.16421}{{\ttfamily 2210.16421}}].

\bibitem{2023ApJ...953..136K}
H.~{Kim}, N.S.~{Kern}, J.N.~{Hewitt}, B.D.~{Nhan}, J.S.~{Dillon}, E.~{de Lera
  Acedo} et~al., \emph{{The Impact of Beam Variations on Power Spectrum
  Estimation for 21 cm Cosmology. II. Mitigation of Foreground Systematics for
  HERA}}, \href{https://doi.org/10.3847/1538-4357/ace35e}{\emph{apj} {\bfseries
  953} (2023) 136} [\href{https://arxiv.org/abs/2307.12826}{{\ttfamily
  2307.12826}}].

\bibitem{2010MNRAS.408.1029L}
A.~{Liu}, M.~{Tegmark}, S.~{Morrison}, A.~{Lutomirski} and M.~{Zaldarriaga},
  \emph{{Precision calibration of radio interferometers using redundant
  baselines}},
  \href{https://doi.org/10.1111/j.1365-2966.2010.17174.x}{\emph{\mnras}
  {\bfseries 408} (2010) 1029}
  [\href{https://arxiv.org/abs/1001.5268}{{\ttfamily 1001.5268}}].

\bibitem{2017PASP..129d5001D}
D.R.~{DeBoer}, A.R.~{Parsons}, J.E.~{Aguirre}, P.~{Alexander}, Z.S.~{Ali},
  A.P.~{Beardsley} et~al., \emph{{Hydrogen Epoch of Reionization Array
  (HERA)}},
  \href{https://doi.org/10.1088/1538-3873/129/974/045001}{\emph{\pasp}
  {\bfseries 129} (2017) 045001}
  [\href{https://arxiv.org/abs/1606.07473}{{\ttfamily 1606.07473}}].

\bibitem{2020MNRAS.499.5840D}
J.S.~{Dillon}, M.~{Lee}, Z.S.~{Ali}, A.R.~{Parsons}, N.~{Orosz},
  C.D.~{Nunhokee} et~al., \emph{{Redundant-baseline calibration of the hydrogen
  epoch of reionization array}},
  \href{https://doi.org/10.1093/mnras/staa3001}{\emph{\mnras} {\bfseries 499}
  (2020) 5840} [\href{https://arxiv.org/abs/2003.08399}{{\ttfamily
  2003.08399}}].

\bibitem{2020ApJ...890..122K}
N.S.~{Kern}, J.S.~{Dillon}, A.R.~{Parsons}, C.L.~{Carilli}, G.~{Bernardi},
  Z.~{Abdurashidova} et~al., \emph{{Absolute Calibration Strategies for the
  Hydrogen Epoch of Reionization Array and Their Impact on the 21 cm Power
  Spectrum}}, \href{https://doi.org/10.3847/1538-4357/ab67bc}{\emph{\apj}
  {\bfseries 890} (2020) 122}
  [\href{https://arxiv.org/abs/1910.12943}{{\ttfamily 1910.12943}}].

\bibitem{2021MNRAS.506.2066C}
S.~{Choudhuri}, P.~{Bull} and H.~{Garsden}, \emph{{Patterns of primary beam
  non-redundancy in close-packed 21 cm array observations}},
  \href{https://doi.org/10.1093/mnras/stab1795}{\emph{\mnras} {\bfseries 506}
  (2021) 2066} [\href{https://arxiv.org/abs/2101.02684}{{\ttfamily
  2101.02684}}].

\bibitem{2024MNRAS.534.2475C}
A.~{Chokshi}, N.~{Barry}, J.L.B.~{Line}, C.H.~{Jordan}, B.~{Pindor} and
  R.L.~{Webster}, \emph{{The necessity of individually validated beam models
  for an interferometric epoch of reionization detection}},
  \href{https://doi.org/10.1093/mnras/stae2264}{\emph{\mnras} {\bfseries 534}
  (2024) 2475} [\href{https://arxiv.org/abs/2409.19875}{{\ttfamily
  2409.19875}}].

\bibitem{2019MNRAS.490..243C}
A.~{Chakraborty}, N.~{Roy}, A.~{Datta}, S.~{Choudhuri}, K.K.~{Datta},
  P.~{Dutta} et~al., \emph{{Detailed study of ELAIS N1 field with the uGMRT -
  II. Source properties and spectral variation of foreground power spectrum
  from 300-500 MHz observations}},
  \href{https://doi.org/10.1093/mnras/stz2533}{\emph{\mnras} {\bfseries 490}
  (2019) 243} [\href{https://arxiv.org/abs/1908.10380}{{\ttfamily
  1908.10380}}].

\bibitem{2000MNRAS.316..749O}
S.~{Oliver}, M.~{Rowan-Robinson}, D.M.~{Alexander}, O.~{Almaini},
  M.~{Balcells}, A.C.~{Baker} et~al., \emph{{The European Large Area ISO Survey
  - I. Goals, definition and observations}},
  \href{https://doi.org/10.1046/j.1365-8711.2000.03550.x}{\emph{\mnras}
  {\bfseries 316} (2000) 749}
  [\href{https://arxiv.org/abs/astro-ph/0003263}{{\ttfamily
  astro-ph/0003263}}].

\bibitem{2021ApJ...907L...7C}
A.~{Chakraborty}, A.~{Datta}, N.~{Roy}, S.~{Bharadwaj}, T.R.~{Choudhury},
  K.K.~{Datta} et~al., \emph{{First Multi-redshift Limits on Post-Epoch of
  Reionization 21 cm Signal from z = 1.96-3.58 Using uGMRT}},
  \href{https://doi.org/10.3847/2041-8213/abd17a}{\emph{\apjl} {\bfseries 907}
  (2021) L7} [\href{https://arxiv.org/abs/2012.04674}{{\ttfamily 2012.04674}}].

\bibitem{2023MNRAS.525.3439E}
K.M.A.~{Elahi}, S.~{Bharadwaj}, S.~{Pal}, A.~{Ghosh}, S.S.~{Ali},
  S.~{Choudhuri} et~al., \emph{{Towards 21-cm intensity mapping at z = 2.28
  with uGMRT using the tapered gridded estimator - III. Foreground removal}},
  \href{https://doi.org/10.1093/mnras/stad2495}{\emph{\mnras} {\bfseries 525}
  (2023) 3439} [\href{https://arxiv.org/abs/2308.08284}{{\ttfamily
  2308.08284}}].

\bibitem{2024MNRAS.529.3372E}
K.M.A.~{Elahi}, S.~{Bharadwaj}, S.~{Pal}, A.~{Ghosh}, S.S.~{Ali},
  S.~{Choudhuri} et~al., \emph{{Towards 21-cm intensity mapping at z = 2.28
  with uGMRT using the tapered gridded estimator - IV. Wide-band analysis}},
  \href{https://doi.org/10.1093/mnras/stae740}{\emph{\mnras} {\bfseries 529}
  (2024) 3372} [\href{https://arxiv.org/abs/2403.06736}{{\ttfamily
  2403.06736}}].

\bibitem{2016PASA...33...19T}
C.M.~{Trott} and R.B.~{Wayth}, \emph{{Spectral Calibration Requirements of
  Radio Interferometers for Epoch of Reionisation Science with the SKA}},
  \href{https://doi.org/10.1017/pasa.2016.18}{\emph{\pasa} {\bfseries 33}
  (2016) e019} [\href{https://arxiv.org/abs/1604.03273}{{\ttfamily
  1604.03273}}].

\bibitem{2025arXiv251025886B}
E.~{Beohar}, A.~{Datta}, A.~{Tripathi}, S.K.~{Pal} and R.~{Sagar},
  \emph{{Mitigating gain calibration errors from EoR observations with SKA1-Low
  AA*}}, \href{https://doi.org/10.48550/arXiv.2510.25886}{\emph{arXiv e-prints}
  (2025) arXiv:2510.25886} [\href{https://arxiv.org/abs/2510.25886}{{\ttfamily
  2510.25886}}].

\bibitem{2025arXiv250609533B}
A.~{Bonaldi}, P.~{Hartley}, S.~{Purser}, O.~{Bait}, E.~{Lee}, R.~{Braun}
  et~al., \emph{{SKA-Low simulations for a cosmic dawn/epoch of reionisation
  deep field}}, \href{https://doi.org/10.48550/arXiv.2506.09533}{\emph{arXiv
  e-prints} (2025) arXiv:2506.09533}
  [\href{https://arxiv.org/abs/2506.09533}{{\ttfamily 2506.09533}}].

\bibitem{2020MNRAS.495.3683K}
J.~{Kumar}, P.~{Dutta} and N.~{Roy}, \emph{{Calibration requirements for epoch
  of reionization 21-cm signal observations - I. Effect of time-correlated
  gains}}, \href{https://doi.org/10.1093/mnras/staa1371}{\emph{\mnras}
  {\bfseries 495} (2020) 3683}
  [\href{https://arxiv.org/abs/2005.07126}{{\ttfamily 2005.07126}}].

\bibitem{2014MNRAS.445.4351C}
S.~{Choudhuri}, S.~{Bharadwaj}, A.~{Ghosh} and S.S.~{Ali},
  \emph{{Visibility-based angular power spectrum estimation in low-frequency
  radio interferometric observations}},
  \href{https://doi.org/10.1093/mnras/stu2027}{\emph{\mnras} {\bfseries 445}
  (2014) 4351} [\href{https://arxiv.org/abs/1409.7789}{{\ttfamily 1409.7789}}].

\bibitem{2022MNRAS.512..186K}
J.~{Kumar}, P.~{Dutta}, S.~{Choudhuri} and N.~{Roy}, \emph{{Calibration
  requirements for Epoch of Reionization 21-cm signal observations - II.
  Analytical estimation of the bias and variance with time-correlated residual
  gains}}, \href{https://doi.org/10.1093/mnras/stac499}{\emph{\mnras}
  {\bfseries 512} (2022) 186}
  [\href{https://arxiv.org/abs/2202.09589}{{\ttfamily 2202.09589}}].

\bibitem{2024JCAP...05..068G}
S.~{Gayen}, R.~{Sagar}, S.~{Mangla}, P.~{Dutta}, N.~{Roy}, A.~{Chakraborty}
  et~al., \emph{{Calibration requirement for Epoch of Reionization 21-cm signal
  observation. Part III. Bias and variance in uGMRT ELAIS-N1 field power
  spectrum}}, \href{https://doi.org/10.1088/1475-7516/2024/05/068}{\emph{\jcap}
  {\bfseries 2024} (2024) 068}
  [\href{https://arxiv.org/abs/2402.18306}{{\ttfamily 2402.18306}}].

\bibitem{2025JCAP...07..024G}
S.~{Gayen}, J.~{Kumar}, P.~{Dutta}, K.M.A.~{Elahi}, S.~{Choudhuri} and
  N.~{Roy}, \emph{{Calibration requirements for epoch of reionization 21-cm
  signal observations. Part IV. Bias and variance with time and frequency
  correlated residual gains}},
  \href{https://doi.org/10.1088/1475-7516/2025/07/024}{\emph{jcap} {\bfseries
  2025} (2025) 024} [\href{https://arxiv.org/abs/2503.23825}{{\ttfamily
  2503.23825}}].

\bibitem{1996A&AS..117..137H}
J.P.~{Hamaker}, J.D.~{Bregman} and R.J.~{Sault}, \emph{{Understanding radio
  polarimetry. I. Mathematical foundations.}}, {\emph{\aaps} {\bfseries 117}
  (1996) 137}.

\bibitem{SynthesisImaging}
G.B.~{Taylor}, C.L.~{Carilli} and R.A.~{Perley}, \emph{{Synthesis Imaging in
  Radio Astronomy II}}, vol.~180 (1999).

\bibitem{1986isra.book.....T}
A.R.~{Thompson}, J.M.~{Moran} and G.W.~{Swenson}, \emph{{Interferometry and
  synthesis in radio astronomy}} (1986).

\bibitem{2011A&A...527A.106S}
O.M.~{Smirnov}, \emph{{Revisiting the radio interferometer measurement
  equation. I. A full-sky Jones formalism}},
  \href{https://doi.org/10.1051/0004-6361/201016082}{\emph{\aap} {\bfseries
  527} (2011) A106} [\href{https://arxiv.org/abs/1101.1764}{{\ttfamily
  1101.1764}}].

\bibitem{2013MNRAS.435..597K}
S.~{Kazemi} and S.~{Yatawatta}, \emph{{Robust radio interferometric calibration
  using the t-distribution}},
  \href{https://doi.org/10.1093/mnras/stt1347}{\emph{\mnras} {\bfseries 435}
  (2013) 597} [\href{https://arxiv.org/abs/1307.5040}{{\ttfamily 1307.5040}}].

\bibitem{2017isra.book.....T}
A.R.~{Thompson}, J.M.~{Moran} and G.W.~{Swenson}, Jr., \emph{{Interferometry
  and Synthesis in Radio Astronomy, 3rd Edition}} (2017),
  \href{https://doi.org/10.1007/978-3-319-44431-4}{10.1007/978-3-319-44431-4}.

\bibitem{Pramanick_2025}
S.~Pramanick, S.~Bharadwaj, K.M.A.~Elahi and R.~Mondal, \emph{The evolving
  power spectrum for the light cone epoch of reionization 21 cm signal},
  \href{https://doi.org/10.3847/1538-4357/adc678}{\emph{The Astrophysical
  Journal} {\bfseries 984} (2025) 105}.

\bibitem{2021MNRAS.501.3378P}
S.~{Pal}, S.~{Bharadwaj}, A.~{Ghosh} and S.~{Choudhuri}, \emph{{Demonstrating
  the Tapered Gridded Estimator (TGE) for the cosmological H I 21-cm power
  spectrum using 150-MHz GMRT observations}},
  \href{https://doi.org/10.1093/mnras/staa3831}{\emph{\mnras} {\bfseries 501}
  (2021) 3378} [\href{https://arxiv.org/abs/2012.04998}{{\ttfamily
  2012.04998}}].

\bibitem{2007ASPC..376..127M}
J.P.~{McMullin}, B.~{Waters}, D.~{Schiebel}, W.~{Young} and K.~{Golap},
  \emph{{CASA Architecture and Applications}},  in \emph{Astronomical Data
  Analysis Software and Systems XVI}, R.A.~{Shaw}, F.~{Hill} and D.J.~{Bell},
  eds., vol.~376 of \emph{Astronomical Society of the Pacific Conference
  Series}, p.~127, Oct., 2007.

\end{thebibliography}\endgroup

\pagebreak

\appendix
\section{Supplementary residual-gain statistics}

Summary statistics of the residual gain  errors for all scans across all observing days are shown here. Each plot contains four rows corresponding to the real and imaginary components of the two polarisations. Columns separated by black dashed lines represent different scans. The standard deviation, skewness, and kurtosis are shown for each scan, with error bars indicating antenna-to-antenna variation.
\begin{figure}[h!]
    \centering
    \includegraphics[width=0.725\textwidth]{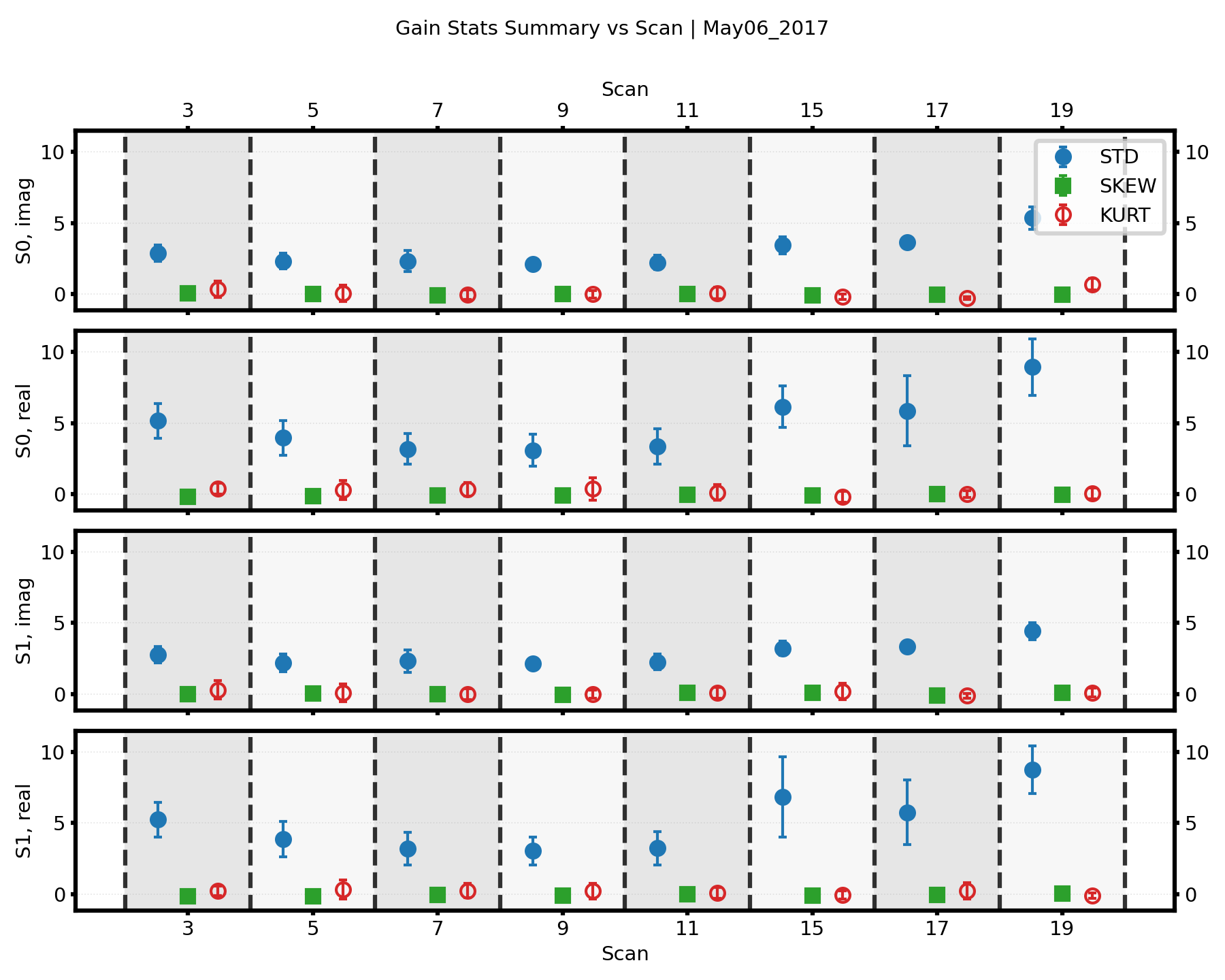}
    \label{fig:gsumm06}
\end{figure}
\begin{figure}[h!]
    \centering
    \includegraphics[width=0.725\textwidth]{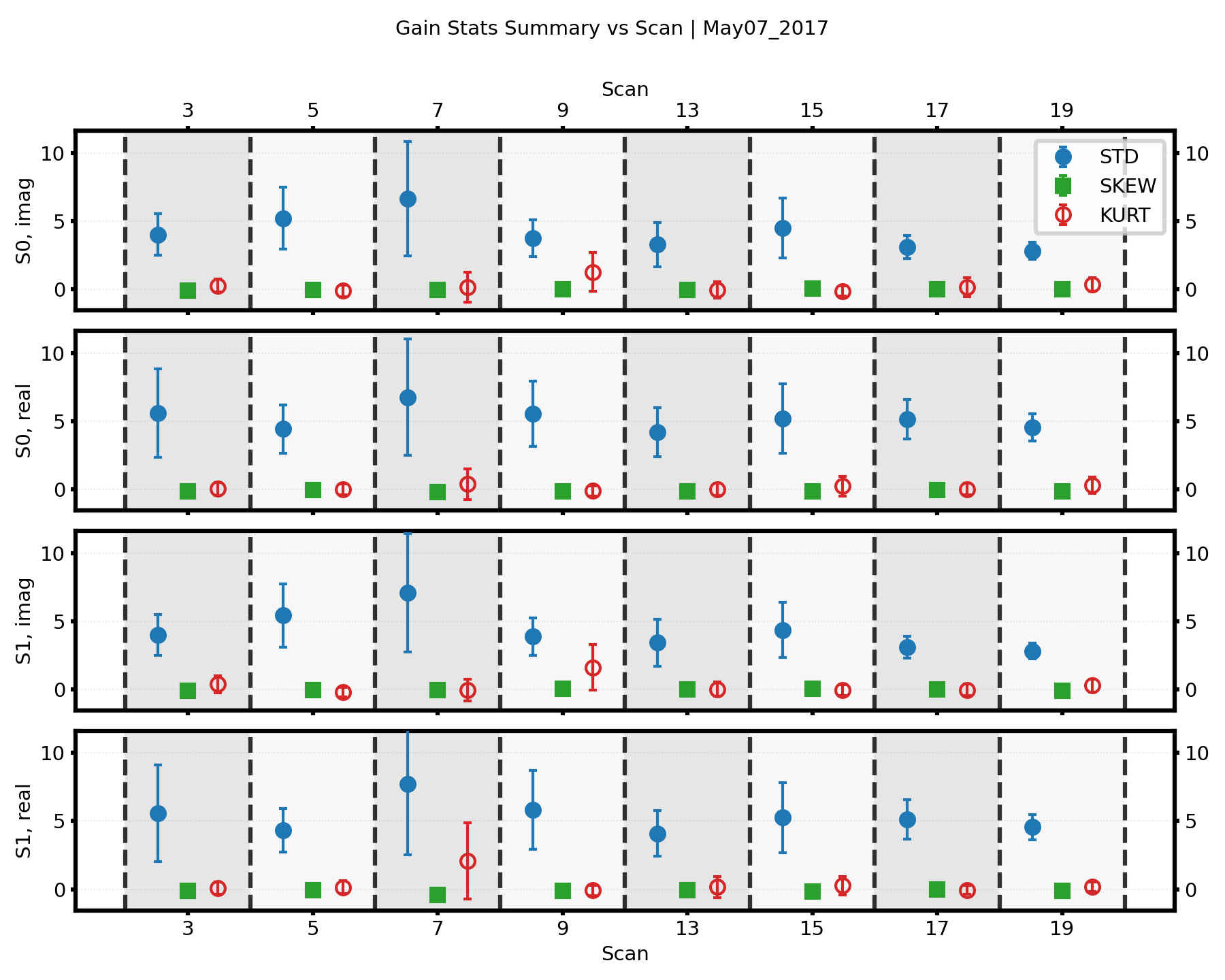}
    \label{fig:gsumm07}
\end{figure}
\begin{figure}[h!]
    \centering
    \includegraphics[width=0.75\textwidth]{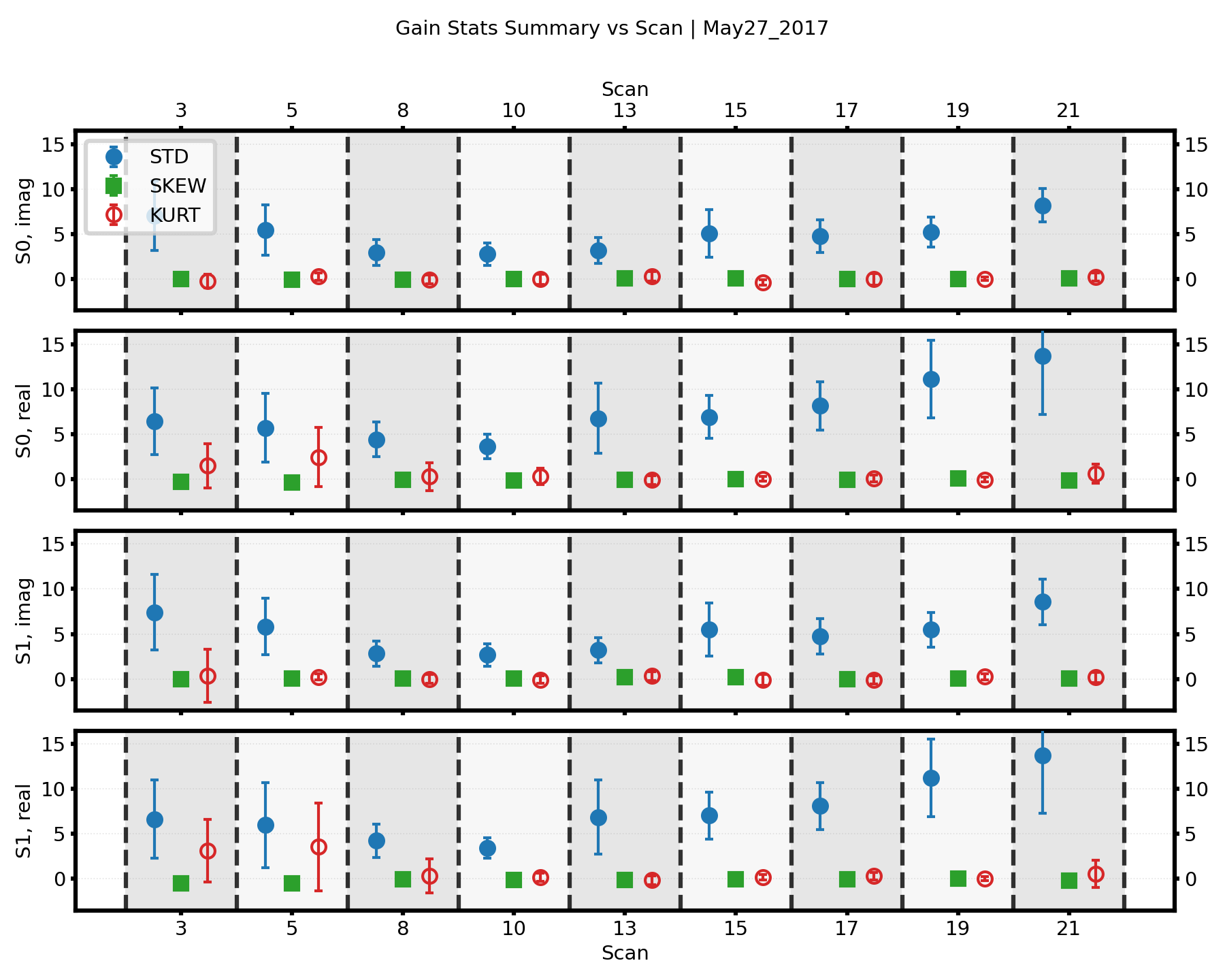}
    \label{fig:gsumm27}
\end{figure}

\end{document}